\documentclass[namedreferences]{SolarPhysics}
\usepackage[optionalrh,natbib]{spr-sola-addons} 
\usepackage{graphicx,rotating}        
\usepackage{amssymb}        
\usepackage{color}           
\usepackage{url}             

\def\bxi{\mbox{\boldmath$\xi$}}

\def\unitx{\mbox{\boldmath$\hat{x}$}}
\def\unitz{\mbox{\boldmath$\hat{z}$}}

\def\unity{\mbox{\boldmath$\hat{y}$}}

\def\bnabla{\mbox{\boldmath$\nabla$}}

\def\bnabla{\mbox{\boldmath$\nabla$}}

\newcommand{\bA}{\mathbf{A}}
\newcommand{\daz}{\partial_z a}

\newcommand{\ddz}{\partial_z}

\newcommand{\bk}{\mathbf{k}}

\newcommand{\rev}[1]{{\textcolor{black}{#1}}}
\newcommand{\revtwo}[1]{{\tt \bf  \textcolor{red}{#1}}}

\begin{document}

\begin{opening}

\title{\rev{Constructing and Characterising Solar Structure Models for Computational Helioseismology}}

\author{H.~\surname{Schunker}$^{1}$ \sep
        R.H.~\surname{Cameron}$^{1}$ \sep
        L.~\surname{Gizon}$^{*  \, 1,2}$ \sep
         H.~\surname{Moradi}$^{1}$
       }
       
\runningauthor{H. Schunker\,\textit{et~al}.}
\runningtitle{Seismic Models for Numerical Helioseismology}

  \institute{$^{1}$ Max-Planck-Institut f\"ur Sonnensystemforschung, Max-Planck Str. 2, 37191 Katlenburg-Lindau, Germany \\
             $^{2}$ Institut f\"ur Astrophysik, Georg-August-Universit\"at G\"ottingen, 37077 G\"ottingen, Germany \\
                     $^*$email: \url{gizon@mps.mpg.de} 
             }

\date{Received ; accepted }

\begin{abstract}
In local helioseismology, numerical simulations of wave propagation are
useful to model the interaction of solar waves with perturbations to a
background solar model. However, the solution to the linearised equations of
motion include convective modes that can swamp the helioseismic waves we are
interested in. In this paper, we construct background solar models that are
stable against convection, by modifying the vertical pressure gradient of
Model~S (Christensen-Dalsgaard \textit{et al}., 1996, \textit{Science}, \textbf{272}, 1286) relinquishing hydrostatic
equilibrium. However, the stabilisation affects the eigenmodes that we wish
to remain as \rev{close to Model~S} as possible. In a bid to recover the Model~S
eigenmodes, we choose to make additional corrections to the sound speed of
Model~S before stabilisation. No stabilised model can be perfectly
solar-like, so we present three stabilised models with slightly different
eigenmodes. The models are appropriate to study the \textit{f} and \textit{p}$_1$ to \textit{p}$_4$ 
modes with spherical harmonic degrees in the range from 400 to 900. Background model CSM has a
modified pressure gradient for stabilisation and has eigenfrequencies within
2\% of Model~S. Model CSM\_A has an additional 10\% increase in sound speed
in the top 1~Mm resulting in  eigenfrequencies within 2\% of Model~S and
eigenfunctions that are, in comparison with CSM, closest to those of Model~S.
Model CSM\_B has a 3\% decrease in sound speed in the top 5~Mm
resulting in eigenfrequencies within 1\% of Model~S and eigenfunctions that
are only marginally adversely affected. These models are useful to study the
interaction of solar waves with embedded three-dimensional heterogeneities,
such as convective flows and model sunspots. We have also calculated the
response of the stabilised models to excitation by random near-surface
sources, using simulations of the propagation of linear waves. We find that
the simulated power spectra of wave motion are in good agreement with an
observed SOHO/MDI power spectrum. Overall, our convectively stabilised
background models provide a good basis for quantitative numerical local
helioseismology. The models are available for download from
\tt{http://www.mps.mpg.de/projects/seismo/NA4/}.
\end{abstract}


\end{opening}

\section{Introduction}\label{Introduction}

Numerical simulations are an important tool to study the effects of surface and subsurface solar structures (sunspots, flows, \textit{etc}.) on solar oscillations. 
Since the wave amplitudes are small compared to the unperturbed background, the equations of motion can be  linearised about a background solar model containing the solar structure being studied.
One requirement of linear simulations is that the medium through which the waves propagate must be stable against convection to prevent unstable modes, which grow exponentially and quickly dominate the solution. A commonly used approach is to consider polytropic background models which are convectively stable by construction \citep[\textit{\textit{e.g.}}][]{Cally1993}. 
However, the Sun is not a polytrope.

This work is motivated to satisfy the need to have convectively stable background models \rev{with eigenmodes similar to those of Model~S \citep{JCD1996}.
We note that Model~S is not a perfect model of the Sun, however it has the advantage that it has been extensively tested and used in helioseismology.
}

This article is divided into the following sections:
Section~2 specifies the problem: the geometry, the equations of motion, the wave attenuation model, boundary conditions, and the condition for stability. 
Section~3 outlines the strategy for constructing the models and measuring the eigenfrequencies and eigenfunctions. 
Sections~4 through to 7 give a detailed description and characterisation of the eigenmodes of each of the background models that we obtain.
In Section~8 we implement a model of random wave excitation in the Semi-spectral Linear MHD (\textsf{SLiM}) code \citep{Cameron2007} and compute the azimuthally averaged power spectra for CSM\_A and CSM\_B. 
The power spectra are compared to an observed power spectrum from the \textit{Michelson Doppler Imager} onboard the \textit{Solar and Heliospheric Observatory} (SOHO/MDI) \citep{Scherrer1995}. 
We conclude with a short discussion of the models and their foreseen uses.

\section{Specifications of the Problem}

\subsection{Geometry}\label{geom}
In this work we are interested in modelling a relatively small portion of the Sun near the solar surface which extends from 25~Mm below the surface to $2.5$~Mm above and $145.77$~Mm in each of the horizontal directions. 
We define the height [$z$] to be negative below the surface and positive above, with $z=0$ given by Model~S \citep{JCD1996}.
The region is large enough that we can study high-degree low-order ($n \leq 4$) modes. 
Relative to the entire spherical Sun, however, the size of the region is small. 
Therefore, in the horizontal direction we can use Cartesian geometry, rather than spherical, so that the problem may be solved more efficiently in (horizontal) spectral space. 
We retain the spherical treatment in the radial direction. 
In this approximation, the operators of the problem, where $a$ is any scalar field and $\bA$ is any vector field, defined in Section~\ref{lwe} are given explicitly by 
\begin{eqnarray}
\nabla a & \equiv &  \daz \unitz  + \mathrm{i} k_x \unitx + \mathrm{i}  k_y \unity  \label{eqn:A1} \\
\nabla \cdot \bA & \equiv &  \frac{1}{(z + R_\odot)^2} \ddz [(z + R_\odot)^2 A_z]\unitz + \mathrm{i}  k_x A_x \unitx + \mathrm{i}  k_y A_y \unity 
\label{eqn:A2}
\end{eqnarray}
where the horizontal wave vector is given by $\bk = k_x \unitx + k_y \unity$. 
We note here that $z+R_\odot$ is equal to the radial distance from the centre of the Sun.

\subsection{Linearised Wave Equation}\label{lwe}
We want to solve for \rev{waves propagating through} a solar background model in the absence of a flow or magnetic field. 
\rev{
For adiabatic oscillations} the ideal hydrodynamic equations linearised about an arbitrary, inhomogeneous, background, can be written as \citep[\textit{e.g.},][]{LyndenBell1967}:
\begin{equation}
\rho \partial_t^2 \bxi =  \nabla ( c^2 \rho \nabla \cdot \bxi + \bxi \cdot \nabla p)    -\bnabla \cdot (\rho \bxi)  g \unitz 
\label{eqn:F0}
\end{equation}
where $\bxi(\bk,z,t)$ is the displacement vector, and $c$, $p$, $\rho$, and $g < 0$  are the background  sound speed, pressure,  density, and gravitational acceleration respectively. The operators are specified by Equations~(\ref{eqn:A1}) and (\ref{eqn:A2}).
\rev{Waves in the Sun are attenuated by turbulent convection.
We model this by implementing an attenuation parameter, as described in Section~\ref{secatt},
 into Equation~\ref{eqn:F0} in the following way:
 }
\begin{equation}
\rho (\partial_t + \gamma)^2 \bxi =  \nabla ( c^2 \rho \nabla \cdot \bxi + \bxi \cdot \nabla p)    -\bnabla \cdot (\rho \bxi)  g \unitz .
\label{eqn:F}
\end{equation}

We have modelled the attenuation so that it operates both on the displacement and velocity. This assumes that  turbulence in the Sun redistributes the displacement perturbations throughout the atmosphere without necessarily involving the macroscopic (observable) velocity. This leads us to use $\mathbf{v}= (\partial_t + \gamma)\bxi$ as the observable velocity as in \cite{Cameron2008}. 
 
In this article the  \textsf{SLiM} code is used to solve Equation~(\ref{eqn:F})  \citep{Cameron2007} \rev{for two types of simulations: to propagate wave-packets and to simulate the stochastically excited wave field of the Sun.}
The simulations use 1098 uniformly spaced ($0.025$~Mm) grid points in the vertical direction and  100 modes in each of the horizontal directions.

\subsection{Damping Layers and Wave Attenuation}\label{secatt}
We retain the boundary conditions of \cite{Cameron2007} where the box is periodic in the horizontal direction and the top boundary condition is a free surface (the Lagrangian pressure perturbation is zero).  In addition, at the top and bottom boundaries, ``sponge'' layers are implemented that artificially reduce the energy of the waves  to minimise reflection.
\begin{figure}
\includegraphics[width=0.7\textwidth,angle=90]{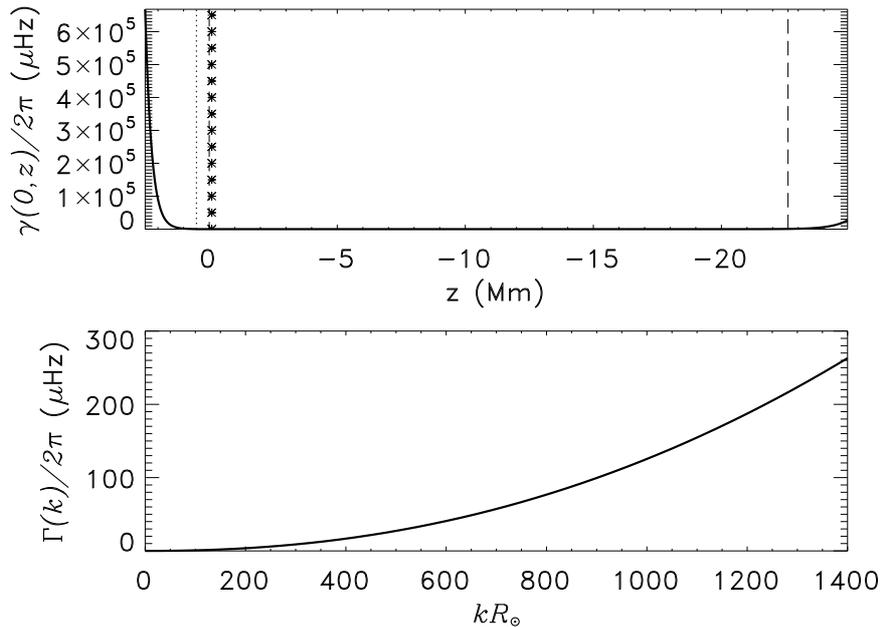}
\caption{The top panel shows the damping $\gamma/2\pi$ (with $\Gamma(k)=0$) as a function of $z$. 
The top damping layer is much stronger than the bottom. 
From left to right, the dotted line is the top of Model~S ($z_t$), the short-dashed line is the surface, the asterisks are the height at which the random sources, $z_{*}$, are implemented and the long-dashed line is the effective bottom of the box ($z_b$). 
The bottom panel is a plot of the attenuation [$\Gamma(k)/2\pi$]  as a function of $kR_\odot$.}
\label{sponge}
\end{figure}

Waves in the Sun are attenuated by granulation and have a finite lifetime.
We model the frequency full width at half maximum of the \textit{f}-mode power using  $\Gamma(k) = \Gamma_* (k/k_*)^{2.2}$, where $\Gamma_* / 2\pi = 100~\mu$Hz  and $k_* = 902/R_\odot$  \citep{Gizon2002}. 
The LHS of Equation~(\ref{eqn:F}) uses $(\partial_t + \gamma)^2 \bxi \approx ( \partial_t^2 + 2 \gamma \partial_t ) \bxi$,  whereas \cite{Gizon2002} use  $(\partial_t + \Gamma)\partial_t \bxi = (\partial_t^2 + \Gamma \partial_t )\bxi$.
Therefore, the attenuation coefficient used in our equation of motion is \textit{half} of that used in \cite{Gizon2002}.
The full form of the damping, $\gamma(k,z)$, shown in Figure~\ref{sponge}, is given by
 \begin{displaymath}
   \frac{\gamma(k,z)}{2\pi} =  \frac{ \Gamma(k)}{4\pi} + \, \\
   \left\{
        \begin{array}{ll|}
            e^{ (z + 0.85~\mathrm{Mm} )/[0.25~\mathrm{Mm}] } \, \mu\mathrm{Hz}  & \textrm{for }0.525<z<2.5\textrm{ Mm}\\	
         e^{ -(z+18.54~\mathrm{Mm})/[0.625~\mathrm{Mm}] } \, \mu\mathrm{Hz}  &  \textrm{for } -25<z<-20\textrm{ Mm.}
     \end{array}
     \right.
\end{displaymath}

The top damping layer introduces a frequency dependence to the eigenmode solutions. High frequency waves have significant energy in the vicinity of the top damping layer and are affected more than the low frequency waves that have less energy at these heights. 
\revtwo{
Any damping layers will affect the eigenfrequencies and lifetimes of the mode, but in this case the lifetimes are predominantly dictated by $\Gamma(k)$.
The parameters for the damping layers were guesses which were shown to empirically damp the reflected waves sufficiently and not noticeably affect the eigenfrequencies or lifetimes of the modes.
 The damping layer parameters are not optimised and other forms have also been found to work \citep[\textit{e.g.}][]{Hanasoge2007}.
}
By using the boundary value problem (BVP) solver in Appendix~\ref{app2} we find that the difference in the eigenfrequencies between having and not having the sponge layers is less than $0.5\%$ for the \textit{f}, \textit{p}$_1$, and \textit{p}$_2$ modes and a little higher for the \textit{p}$_3$ and \textit{p}$_4$ modes (see Appendix~\ref{app3}, Figure~\ref{bvpfig}d). If we adjust the range of the top damping layer to $0.125~\textrm{Mm}<z<2.5$~\textrm{Mm} we see a maximum 0.5\% reduction but only for the  \textit{p}$_4$-modes at high frequencies (see Appendix~\ref{app3}, Figure~\ref{bvpfig}f).

\subsection{Initial Background Model}
We begin with Model~S as our  background model (starting from any other standard solar model would also be possible). Model~S extends to 0.5~Mm above the surface, however our computational domain extends up to 2.5~Mm so that the boundary conditions are sufficiently far from the surface.  We extend Model~S above $z_{t}=0.5$~Mm  in the following way:
\begin{eqnarray}
c_0(z)=c_\mathrm{S}(z_{t}) & \textrm{for } z >  z_t \, ,\\
\rho_0(z)=\rho_\mathrm{S}(z_{t})e^{-(z-z_t)/[0.125~\mathrm{Mm}]} & \textrm{for } z >  z_t \, ,\\
p_0(z)=p_\mathrm{S}(z_{t})e^{-(z-z_t)/[0.15~\mathrm{Mm}]} & \textrm{for } z >  z_t \, ,
\end{eqnarray}
where the subscript ``S'' refers to Model~S, the subscript ``0'' is the extended model. 
The denominators in the exponents are the scale heights of the density and pressure respectively  at $z_t$. 
The only requirement for the extension of the background was that it should not increase the wavespeed since we aim to damp the waves at these heights to minimise reflection.  
Thus, the sound speed was held constant and the pressure and density smoothly extended. 
\rev{The extension is not meant to represent a realistic solar chromosphere and at this height the waves will be artificially damped to prevent reflection.}

\subsection{Conditions for Convective Stability}\label{convecstab}

We want to simulate perturbations superimposed on a background model assuming that the evolution is linear. Part of Model~S, and therefore the extended Model~S described above, is super-adiabatically stratified and convectively unstable. 
This instability is a real property of the Sun resulting in modes that, in a linear calculation, grow exponentially in time and will eventually dominate the solution. 
Therefore, we stabilise the background model against convection to satisfy the condition $\mathrm{d}_z p > c^2 \mathrm{d}_z \rho$. 
We do this by altering the pressure gradient.
The reason for choosing to modify the pressure gradient is that it affects the eigenmodes of the model less than changes to the sound speed and/or density \citep{Cameron2008}. 
We set the pressure gradient in the stabilised model as:
 \begin{displaymath}
   \mathrm{d}_z p =   \\
   \left\{
        \begin{array}{ll}
              \mathrm{max}(c_0^2 \mathrm{d}_z \rho_0, \mathrm{d}_z p_0) & \textrm{for } \, z \le -0.15~\textrm{Mm} \, , \\
               \mathrm{max}(c_0^2 \mathrm{d}_z \rho_0 - \epsilon_1,\mathrm{d}_z p_0) & \textrm{for } \, -0.15~\textrm{Mm} < z <  0.1~\textrm{Mm} \, ,\\
               \mathrm{max}(c_0^2 \mathrm{d}_z \rho_0,\mathrm{d}_z p_0) & \textrm{for } \, 0.1 \le z < 0.325~\textrm{Mm} \, , \\	        
               \mathrm{max}(0.99c_0^2 \mathrm{d}_z \rho_0,\mathrm{d}_z p_0)  &  \textrm{for } z >  0.325~\textrm{Mm}      \, ,
      \label{stab}
     \end{array}
     \right.
\end{displaymath}
where $\epsilon_1=10^{-5}$~cgs (at the surface this is $\approx 0.002 \, c_0^2 \, \mathrm{d}_z \, \rho_0$).
\rev{
This formulation was arrived at by empirically testing the stability of the simulation with small values of $\epsilon_1$.
An additional constraint was that it should also remain stable with an embedded perturbation \citep[\textit{e.g.} a sunspot as in ][]{Cameron2008}.
This was the smallest value that was found to satisfy these conditions.
}
The derivatives, here, are evaluated numerically as
%
\begin{eqnarray*}
\mathrm{d}_z p_0(z_i) & \equiv & p_0(z_i) \ln\left[ p_0(z_{i+1})/p_0(z_{i-1}) \right] / ( z_{i+1} -z_{i-1} ) \\
 \textrm{and} & \, &  \, \\
\mathrm{d}_z \rho_0(z_i) & \equiv & \rho_0(z_i) \ln\left[ \rho_0(z_{i+1})/\rho_0(z_{i-1}) \right] / ( z_{i+1} -z_{i-1} )
\end{eqnarray*}
to achieve a greater numerical accuracy. 
We have tested that this criterion is effective in maintaining stability for simulations for up to ten solar days.\\

The stabilisation forfeits hydrostatic equilibrium and introduces gravity modes into the solution.
The gravity modes all have low frequencies and can easily be excluded from any subsequent analyses.
The lack of hydrostatic equilibrium is likely to be more consequential.
\rev{
There are different formulations of the oscillation equations, those that incorporate the assumption of hydrostatic equilibrium and those that do not. 
We stress that everything presented in this article applies to the formulation presented in Equation~\ref{eqn:F} which was derived from the equations of continuity, energy and motion, respectively
\begin{eqnarray*}
\partial_t \rho^\prime &=& - \nabla \cdot (\rho \partial_t \bxi), \\
\partial_t p^\prime &=& c^2 ( \partial_t \rho^\prime + \partial_t (\bxi \cdot \nabla \rho)) - \partial_t (\bxi \cdot \nabla p) \, \, \, \, \mathrm{and} \\
\rho ( \partial_t + \gamma)^2 \bxi &=& - \nabla p^\prime + \rho^\prime g \unitz ,
\end{eqnarray*}
(where the primed quantities are the perturbations), {\textit{without}} assuming hydrostatic equilibrium. 
Also, the implications for seismic reciprocity \citep{Dahlen1998} have not been explored and may be important. }


\section{Strategy Outline}\label{secstrategy}

Now that we have set out the problem, we outline the strategy involved in developing the convectively stable background models presented in this article. 
 It is described as follows: 
\begin{itemize}
\item Begin with Extended Solar Model~S.
\item Convectively stabilise it by changing $\mathrm{d}_zp$ as described in Section~\ref{convecstab}. This results in CSM.
\item Compare the eigenfrequencies and eigenfunctions to those of Model~S. 
\item We find that the eigenfunctions near the surface, \rev{where we are most interested in modelling,} are not well matched and the eigenfrequencies have increased. 
\end{itemize}
\rev{We are left with the choice to modify the sound speed and/or the density to try to correct the eigenmodes. Since modifying the density has a large effect on the \textit{f}-mode energy density, we choose to change the sound speed only.} 
Empirically, we found that increasing the sound speed near the surface improves the eigenfunctions:
\begin{itemize}
\item Begin with Extended Solar Model~S.
\item Increase the sound speed in the top 1~Mm by 10\% (Equation~\ref{csa}).
\item Convectively stabilise the model. This results in CSM\_A.
\item Compare the eigenfrequencies and eigenfunctions to those of Model~S.  
\item We find that  the eigenfunctions are a better match with Model~S than CSM and the eigenfrequencies are only slightly over-estimated.
\end{itemize}
\rev{We attempt to correct the eigenfrequencies by introducing a small decrease in sound speed in the top $\approx 5$~Mm, which will reduce the overall speed of the waves and thus reduce the eigenfrequencies}:
\begin{itemize}
\item Begin with Extended Solar Model~S.
\item Take the sound speed profile of CSM\_A and introduce an additional decrease in the sound speed of  3\% in the top $\approx 5$~Mm  (Equation~\ref{csb}).
\item Convectively stabilise the model.
\item  Compare the eigenfrequencies and eigenfunctions to those of Model~S.  
\item We find the eigenfrequencies are \rev{closer to Model~S} and the eigenfunctions are only moderately \rev{further from Model~S} than CSM\_A. This results in CSM\_B.
\end{itemize}
\rev{For a smooth transition, a Gaussian function was selected for the sound speed changes.
The particular parameters were determined by trial-and-error of a few guesses to empirically evaluate how they further affected the eigenmodes. 
The comparisons to the eigenmodes of Model~S were judged by eye. }
We calculated the eigenmodes of the models in two ways. The first used the \textsf{SLiM} numerical simulations (see Appendix~\ref{app1}) and the second used a \textsf{BVP} solver (see Appendix~\ref{app2}).  


As a quantitative measure of the  \rev{difference between eigenfrequencies of Model~S and the featured models, we compute the relative difference} of the real part of the eigenfrequencies (determined from both \textsf{SLiM} and the BVP) to the real part of the Model~S eigenfrequencies, $\omega/\omega_\mathrm{S} -1$.   
These particular Model~S eigenfrequencies were calculated \rev{as in \cite{Birch2004}} using a Cartesian geometry and constant gravity.
\rev{For the modes we are interested in, the geometry and radially dependent gravity affect the eigenfrequencies by no more than 0.5\% (see Appendix~\ref{app3}). }
We measure the \rev{difference between  the eigenfunctions of Model~S and the stabilised model in two ways.} 
The first, by  calculating  the relative difference in area under $\textrm{Re}[v_z\sqrt{\rho}]$ near the surface between Model~S and the respective stabilised model. 
The second, by calculating the difference of the height [$z_p$] of the uppermost peak of $\textrm{Re}[v_z\sqrt{\rho}]$  between Model~S and the respective stabilised model for each eigenmode (Section~\ref{sl}).

\section{Convectively Stable Model (CSM)}\label{seccsm}
\begin{center}
\begin{figure}
\includegraphics[width=0.9\textwidth]{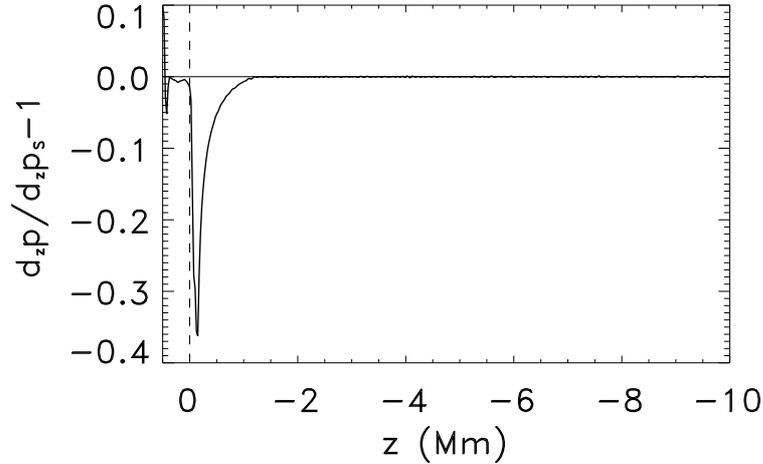}
\caption{The relative difference of the pressure gradient  between CSM and Model~S as a function of height, $z$.}
\label{csm}
\end{figure}
\end{center}

Figure~\ref{csm} shows the relative difference between the stabilised pressure gradient of CSM  and the pressure gradient of Model~S, $\mathrm{d}_z p/ \mathrm{d}_z p_\mathrm{S} -1$, which is as large as 35\% near the surface.
We now discuss the effect this change in the pressure gradient has on the eigenmodes.

\begin{figure}
\hspace{1cm}
\includegraphics[width=0.9\textwidth]{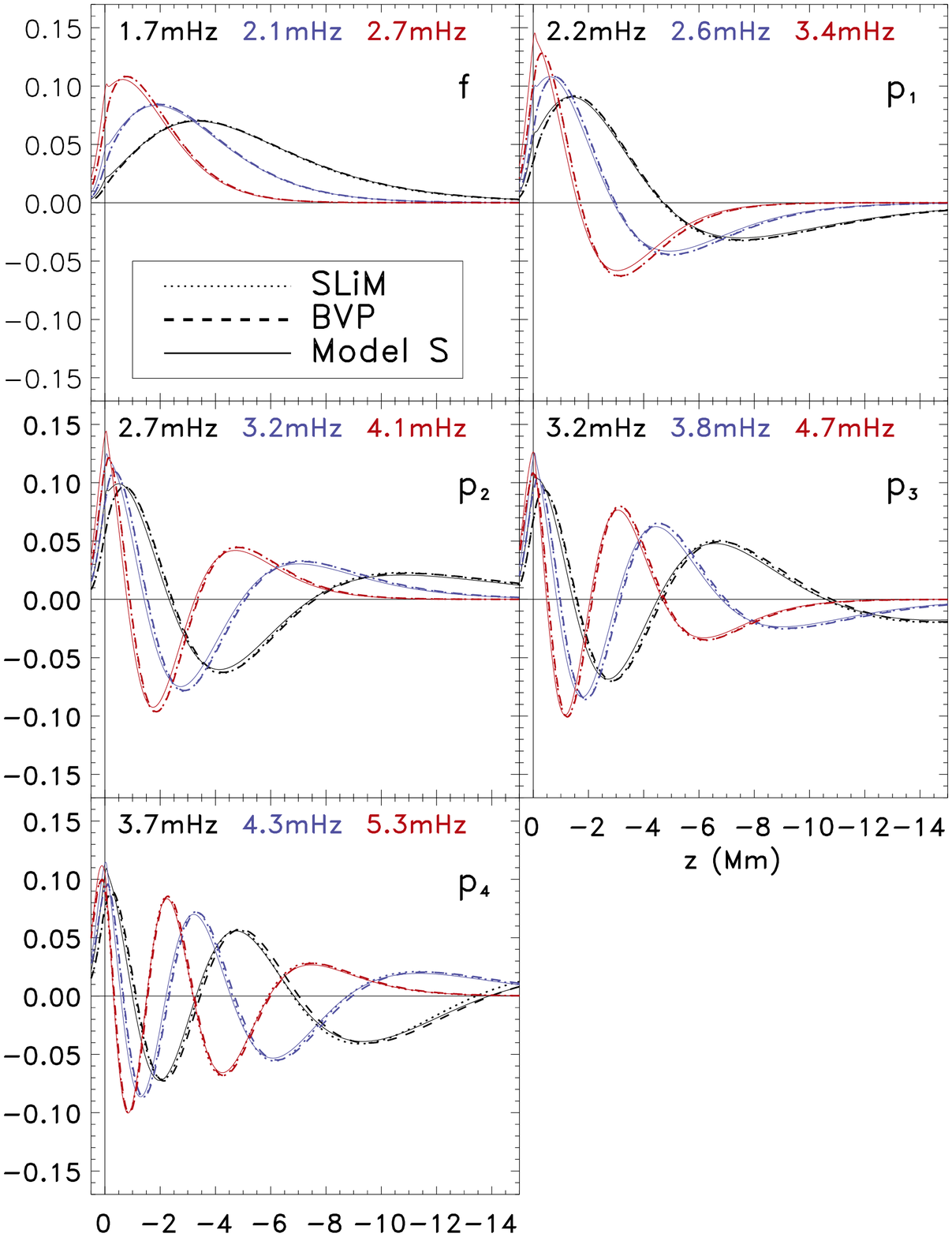}
\vspace{1cm}
\caption{The $z$-dependence of the real component of $v_z\sqrt{\rho}$ for a number of eigenmodes of CSM. 
\rev{The eigenfrequencies for the wavenumbers, $kR_\odot = 270, \, 500, \, 750$, are specified by colour. }
The modes have been normalised so that $v_z$ is real at $0.2$~Mm and have equal integrals.
The dashed curve shows the eigenmodes from the \textsf{BVP} solution, the dotted curve shows the eigenmodes from the \textsf{SLiM} simulations and the solid curve shows the Model~S eigenmodes. 
Each panel corresponds to a different radial order [\textit{f}, \textit{p}$_1$ to \textit{p}$_4$].}
\label{nocs_efunc}
\end{figure}

Figure~\ref{nocs_efunc} shows  $\textrm{Re}[v_z\sqrt{\rho}]$, normalised so that
$\int_{-25\mathrm{Mm}}^{0.5\mathrm{Mm}}\sqrt{(|v_z|^2 + |v_x|^2)\rho} \, \textrm{d}z=1$,  as a function of $z$ for \textit{f}  and \textit{p}$_1$ to \textit{p}$_4$ eigenmodes   from Model~S  and CSM (derived using both \textsf{SLiM} and the \textsf{BVP}). 
Recall that the depth of our domain allows us to study only up to the \textit{p}$_4$ mode.
\rev{The horizontal velocity component of the eigenfunctions, $v_x\sqrt{\rho}$, was found to have a similar agreement with Model~S.}

We observe that the main effect of the stabilisation on  the eigenfunctions is to decrease the amplitude of  $\textrm{Re}[v_z\sqrt{\rho}]$ near the surface.  Since this is where the stabilisation has the greatest effect on the pressure gradient, changes to the eigenfunctions in this region are not unexpected. 

\begin{figure}
\begin{center}
\hspace{1cm}
\includegraphics[width=0.9\textwidth]{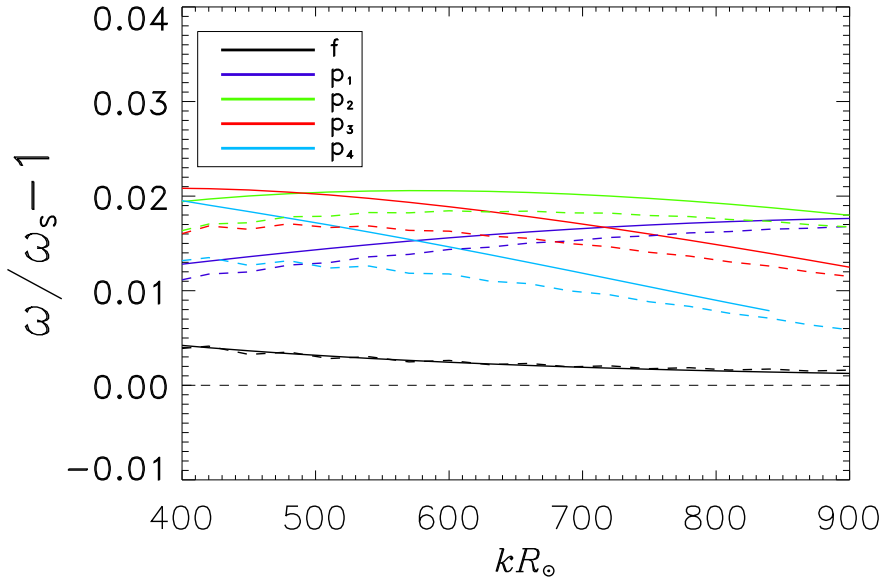}
\caption{The relative difference between the real part of the CSM  eigenfrequencies, $\omega$, and the Model~S eigenfrequencies, $\omega_\mathrm{S}$,  as a function of $kR_\odot$. The solid curves are for the simulated \textsf{SLiM} eigenfrequencies and the dashed curves are for the \textsf{BVP} solutions. The average relative difference of each radial order in this range is within  2\%, \rev{with at most 0.5\% due to the different treatment of gravity and geometry  (see Appendix~\ref{app3}).}}
\label{csmp}
\end{center}
\end{figure}

Figure~\ref{csmp} shows the relative difference of the real part of the eigenfrequencies, $\omega/\omega_\mathrm{S} - 1$, for each radial order as a function of wavenumber. 
The quantitative  average over $400<kR_\odot<900$  shows that the increase in the eigenfrequencies is less than 2\%. 
\rev{The increase in \textit{f}-mode eigenfrequencies compared to Model~S can be attributed to the treatment of gravity and geometry of the operators (see Appendix~\ref{app3}).}
The agreement between Model~S and each of the convectively stable models will be quantified in Section~\ref{sl}.

Since it is a necessity to modify Model~S, and therefore no subsequent model will have exactly the same eigenmodes, we attempt to correct the eigenmodes by modifying the sound speed.
We found a trade-off between having  eigenfunctions or eigenfrequencies \rev{closer to those of Model~S.}
In model CSM\_A (Section~\ref{seccsma}) we attempt to improve the eigenfunctions and
in CSM\_B we try to improve the eigenfrequencies without affecting the eigenfunctions too much (Section~\ref{seccsmb}).


\section{Convectively Stable Model A  (CSM\_A)}\label{seccsma}

\begin{center}
\begin{figure}
\hspace{0.3cm}
\includegraphics[width=0.5\textwidth,angle=90]{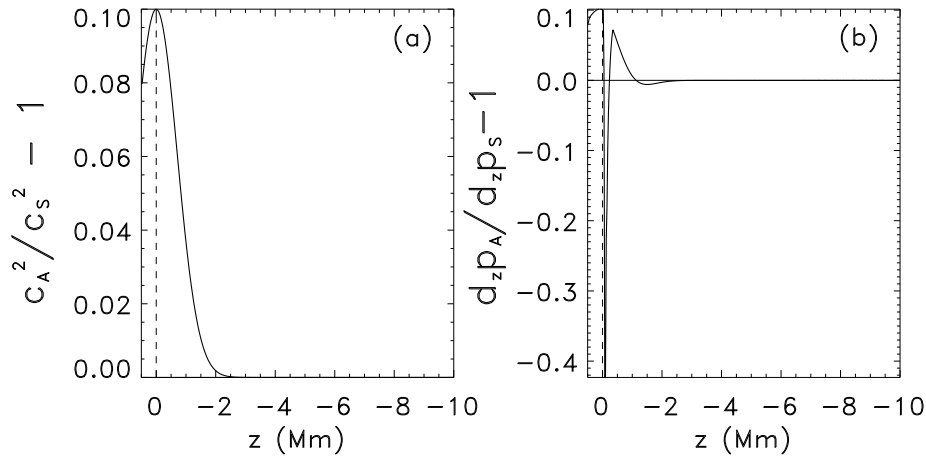}
\vspace{0.5cm}
\caption{ The relative difference   between  CSM\_A and Model~S  of (a) the sound speed squared and (b) the  pressure gradient,  as a function of $z$.}
\label{smsa}
\end{figure}
\end{center}

\begin{figure}
\hspace{1cm}
\includegraphics[width=0.9\textwidth]{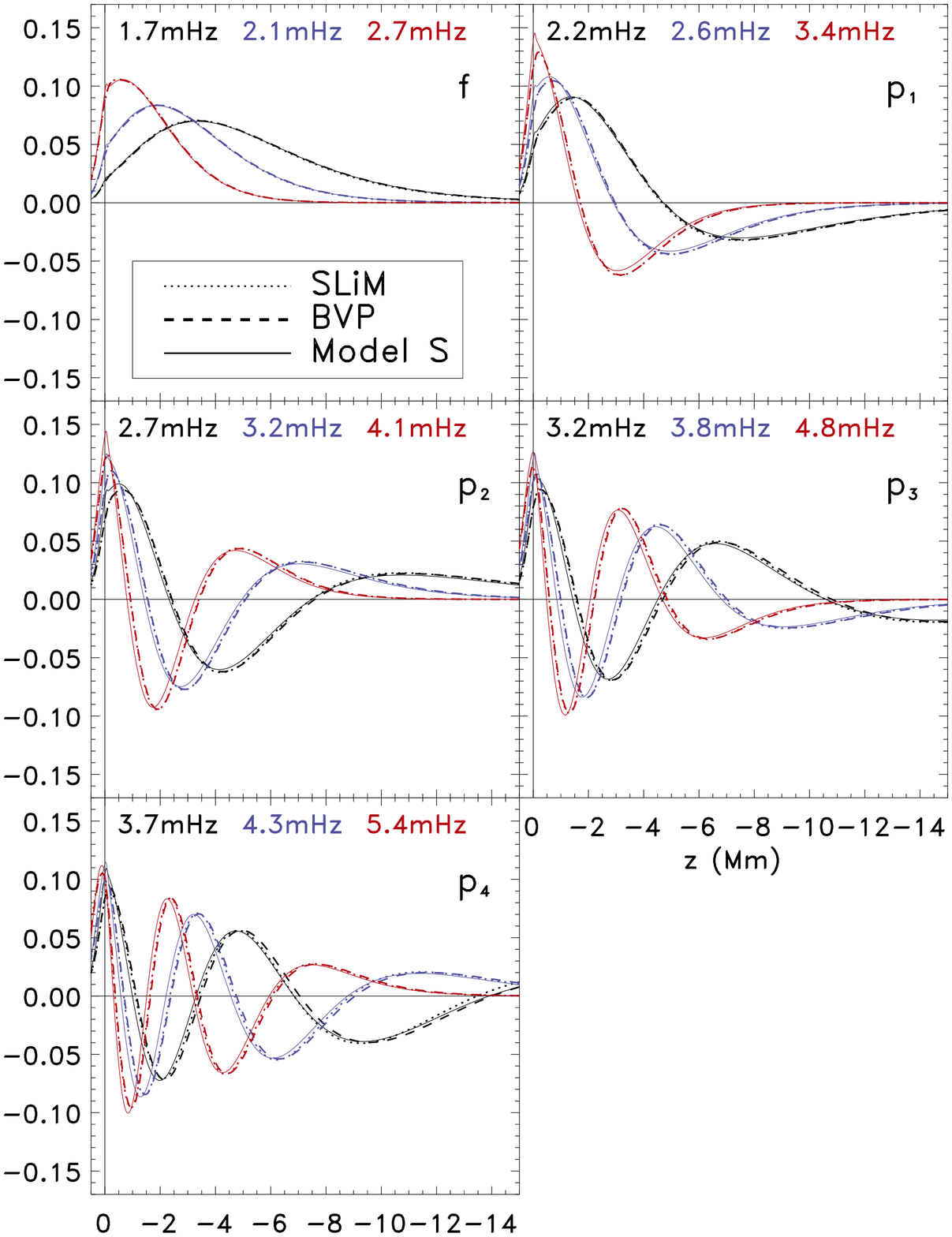}
\vspace{1cm}
\caption{The $z$-dependence of the real component of $v_z\sqrt{\rho}$ for a number of eigenmodes  of CSM\_A. 
\rev{The eigenfrequencies for the wavenumbers, $kR_\odot =  270, \, 500, \, 750$, are specified by colour.  }
The modes have been normalised so that $v_z$ is real at $z=0.2$~Mm and have equal integrals.
The dashed curve shows the eigenmodes from the \textsf{BVP} solution, the dotted curve shows the eigenmodes from the \textsf{SLiM} simulations and the solid curve shows the Model~S eigenmodes. Each panel corresponds to a different radial order [\textit{f}, \textit{p}$_1$ to \textit{p}$_4$].}
\label{smsa_efunc}
\end{figure}

We follow the procedure set out  in Section~\ref{secstrategy}. 
We found that an increase in sound speed improved the match between the eigenfunctions of CSM and Model~S  near the surface.
We chose
\begin{equation}\label{csa}
c_A(z) = c_{\rm 0}(z) \left[ 1 +0.1 \, \mathrm{exp}\left( {-\left(\frac{z}{1.0~\mathrm{Mm}}\right)^2}\right) \right],
\end{equation}
where  the subscript ``A'' indicates  CSM\_A.  
Starting from  Model~S with $c_A$ specifying the sound speed, we then rederived the pressure gradient required for  stability as set out in Section~\ref{convecstab}.  
Figure~\ref{smsa} shows the relative difference between CSM\_A and Model~S of the  sound speed squared and the pressure  gradient  as a function of height.
This change in sound speed was found to raise the height of the uppermost peak of  $\textrm{Re}[v_z \sqrt{\rho}]$. 
Figure~\ref{smsa_efunc} shows $\textrm{Re}[v_z \sqrt{\rho}]$ for various eigenmodes from CSM\_A for each radial order, \textit{f} and \textit{p}$_1$ to \textit{p}$_4$.  Particularly, the \textit{f}-mode eigenfunctions  are close to Model~S. The \textit{p}$_1$ and \textit{p}$_2$ modes are also a better match, especially near the surface. 

\begin{figure}
\begin{center}
\hspace{0.4cm}
\includegraphics[width=0.46\textwidth]{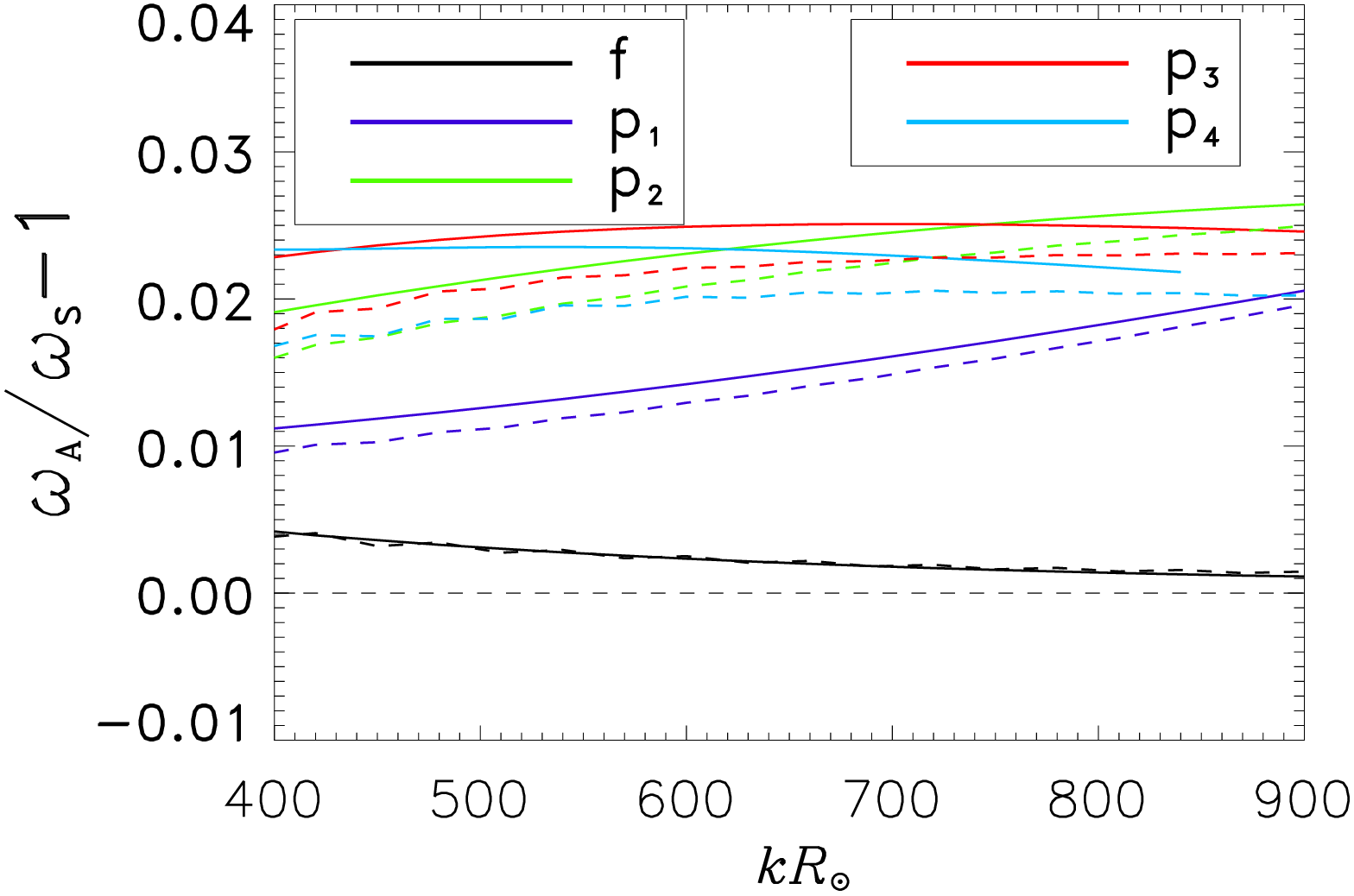}
\hspace{0.1cm}
\includegraphics[width=0.46\textwidth]{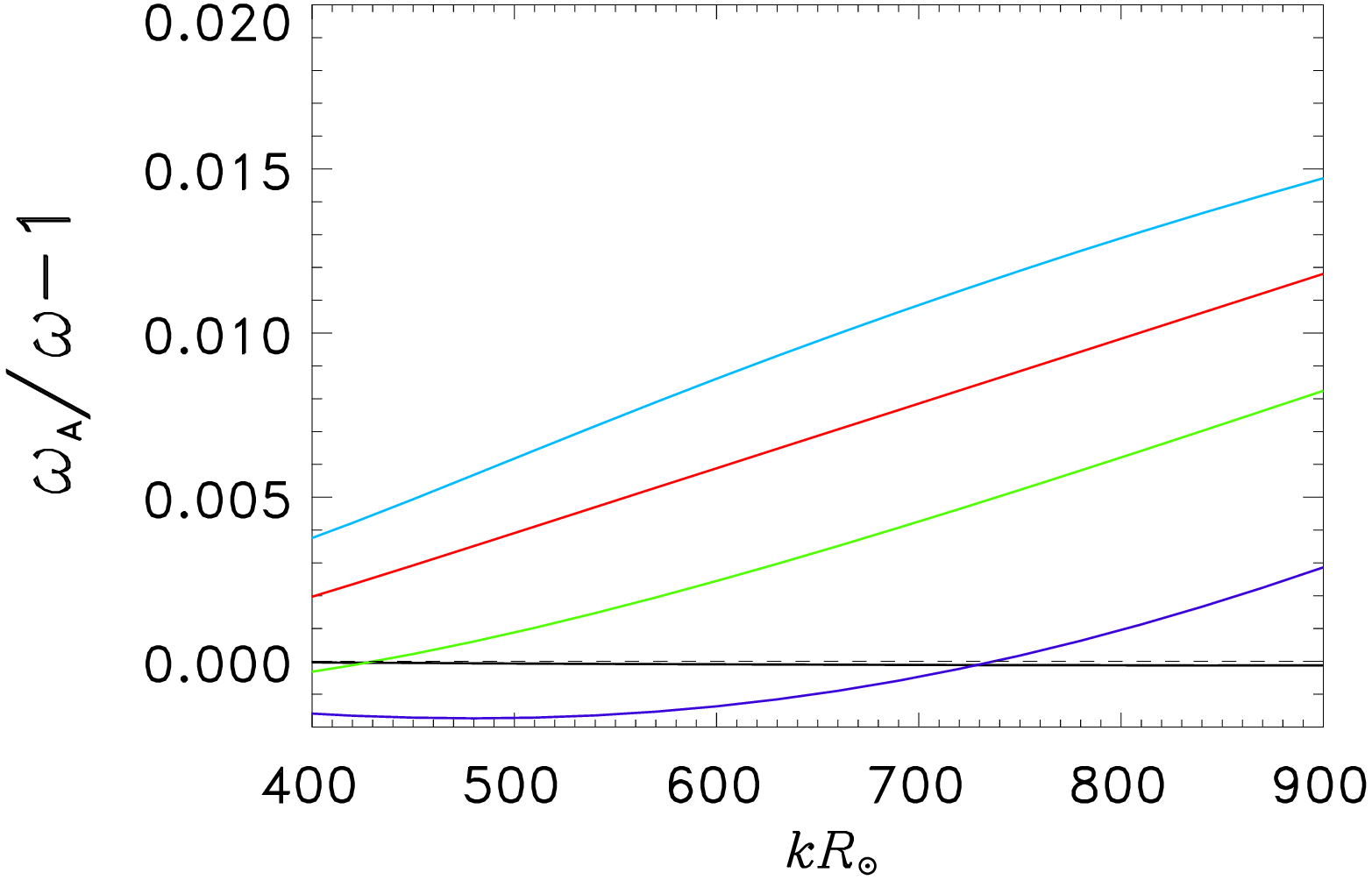}
\caption{Left: the relative difference between the  real part of the CSM\_A eigenfrequencies, $\omega_\mathrm{A}$ and the real part of the Model~S eigenfrequencies, $\omega_\mathrm{S}$  as a function of $kR_\odot$. The solid curves are the differences using the eigenfrequencies calculated from \textsf{SLiM} and the dashed curves from the \textsf{BVP}.  Right: the relative difference between the real part of the CSM\_A eigenfrequencies, $\omega_\mathrm{A}$, and the real part of the CSM eigenfrequencies, $\omega$,   \rev{calculated by \textsf{SLiM}} as a function of $kR_\odot$ brought about by the increase in sound speed.}
\label{efreqa}
\end{center}
\end{figure}

The real parts of the eigenfrequencies, shown in Figure~\ref{efreqa}, are not significantly affected: the average (over $400<kR_\odot<900$) relative difference for each radial order  is still less than 2\% of Model~S values. 
We have constructed  a convectively stable model, CSM\_A, with \rev{eigenfunctions closer to Model~S} than CSM and reasonably \rev{similar} eigenfrequencies. 


\section{Convectively Stable Model B  (CSM\_B)}\label{seccsmb}

\begin{figure}
\begin{center}
\hspace{0.3cm}
\includegraphics[width=0.5\textwidth,angle=90]{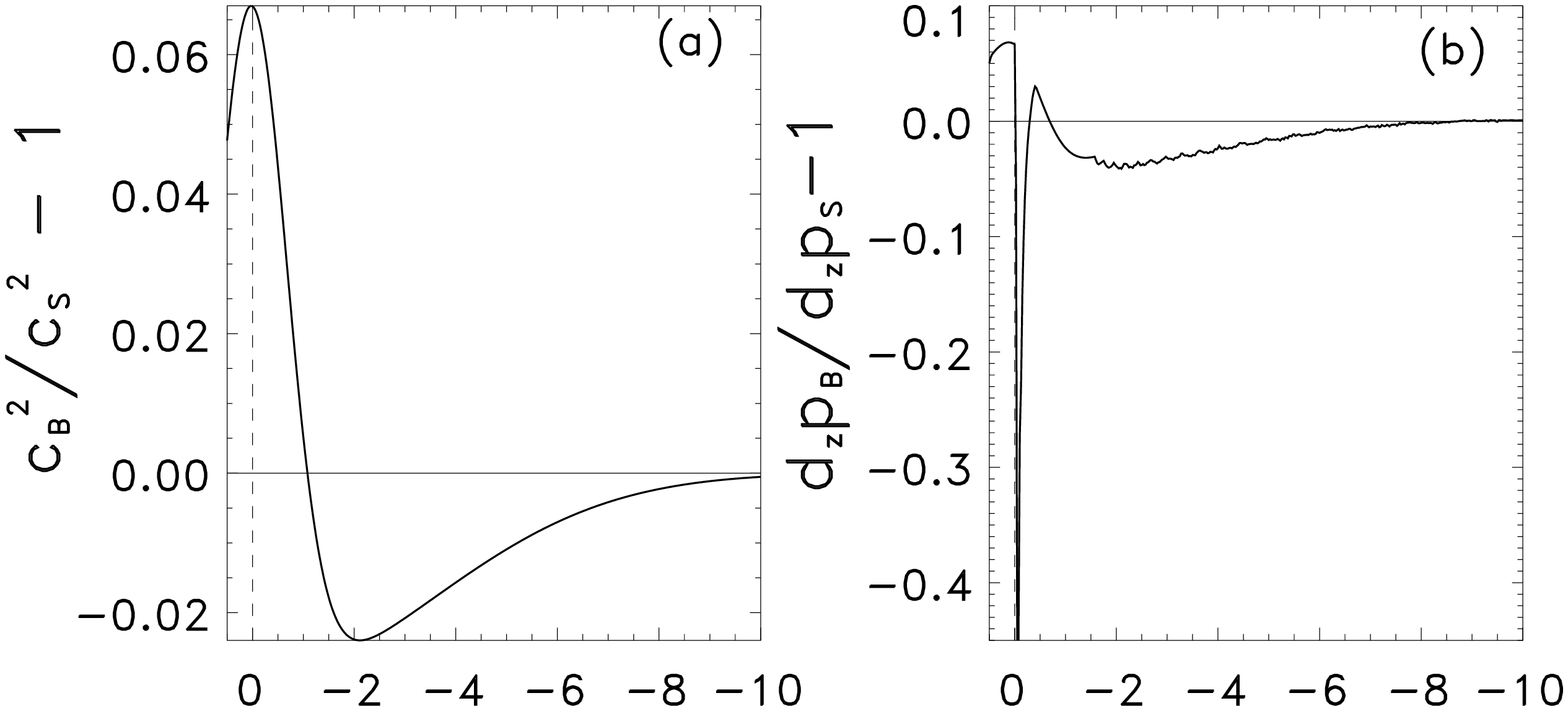}
\caption{The relative difference between CSM\_B and Model~S of (a) the  sound speed squared and (b) the pressure  gradient,  as a function of $z$.}
\label{smsb}
\end{center}
\end{figure}

Starting from Model~S and $c_A$, we constructed a model with \rev{eigenfrequencies closer to Model~S than CSM\_A and reasonable eigenfunctions} (as described in Section~\ref{secstrategy}). 
The eigenfrequencies are related to the phase speed of the wave ($\omega/k$) and so we slowed the waves down by adding a broad reduction in sound speed of CSM\_A.
We chose
\begin{equation}\label{csb}
c_{\rm B}(z)  = c_{\rm A}(z)   \left[ 1 - 0.03 \, \mathrm{exp}\left({-\left(\frac{z}{5.0~\mathrm{Mm}}\right)^2}\right) \right],
\end{equation}
where subscript ``B'' indicates CSM\_B. 
Figure~\ref{smsb} shows the relative difference between CSM\_B and Model~S (a) sound speed squared and (b) pressure gradient as a function of height. 

\begin{figure}
\hspace{0.8cm}
\includegraphics[width=0.9\textwidth]{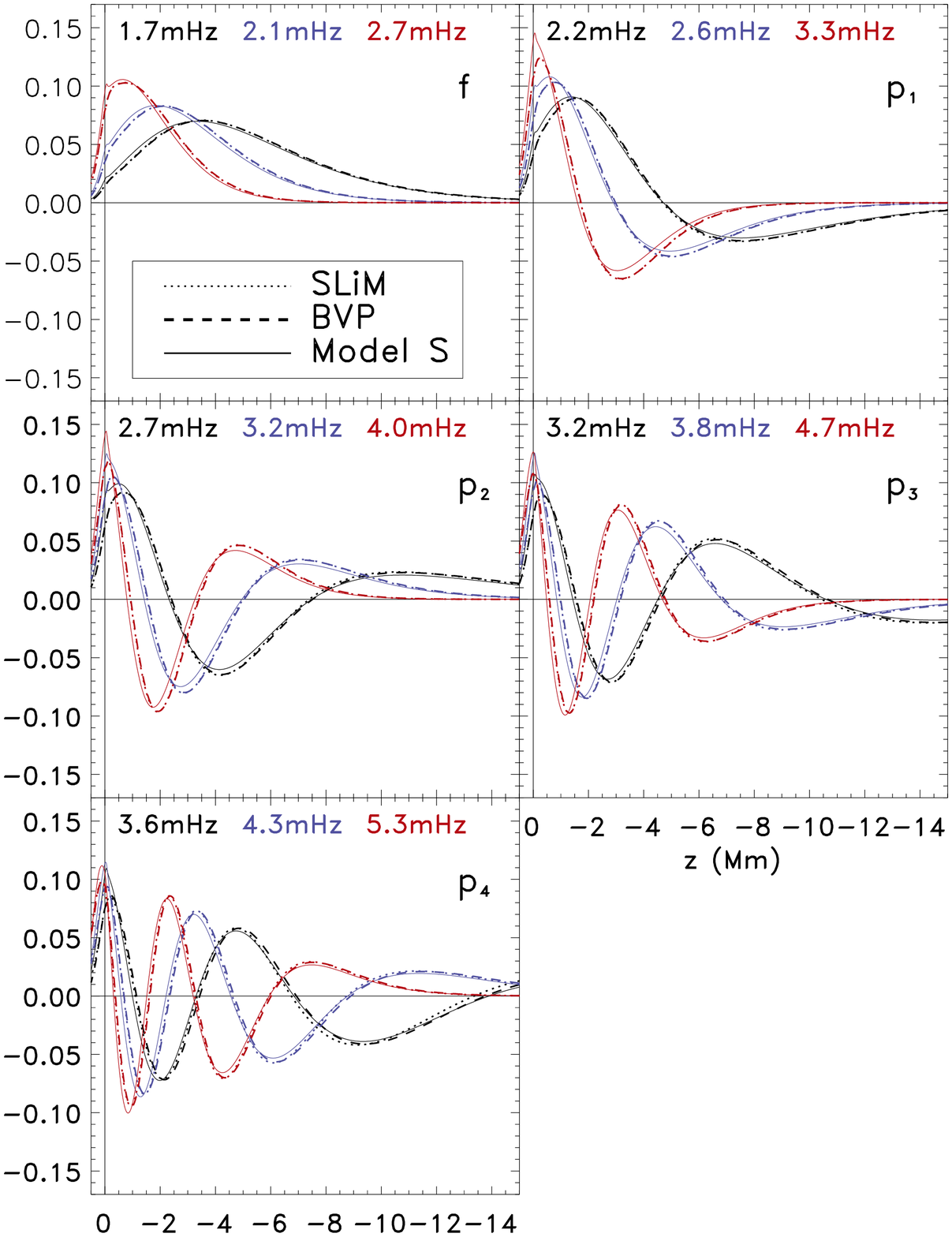}
\vspace{1cm}
\caption{The $z$ dependence of the real component of $v_z\sqrt{\rho}$ for a number of eigenmodes  of CSM\_B. 
\rev{The eigenfrequencies for the wavenumbers, $kR_\odot =  270, \, 500, \, 750$, are specified by colour.  }
The modes have been normalised so that $v_z$ is real at $z=0.2$~Mm and have equal integrals.
The dashed curve shows the eigenmodes from the \textsf{BVP} solution, the dotted curve shows the eigenmodes from the \textsf{SLiM} simulations and the solid curve shows the Model~S eigenmodes. 
Each panel corresponds to a different radial order, \textit{f}, \textit{p}$_1$ to \textit{p}$_4$.}
\label{smsb_efunc}
\end{figure}

\begin{figure}
\begin{center}
\hspace{0.3cm}
\includegraphics[width=0.47\textwidth]{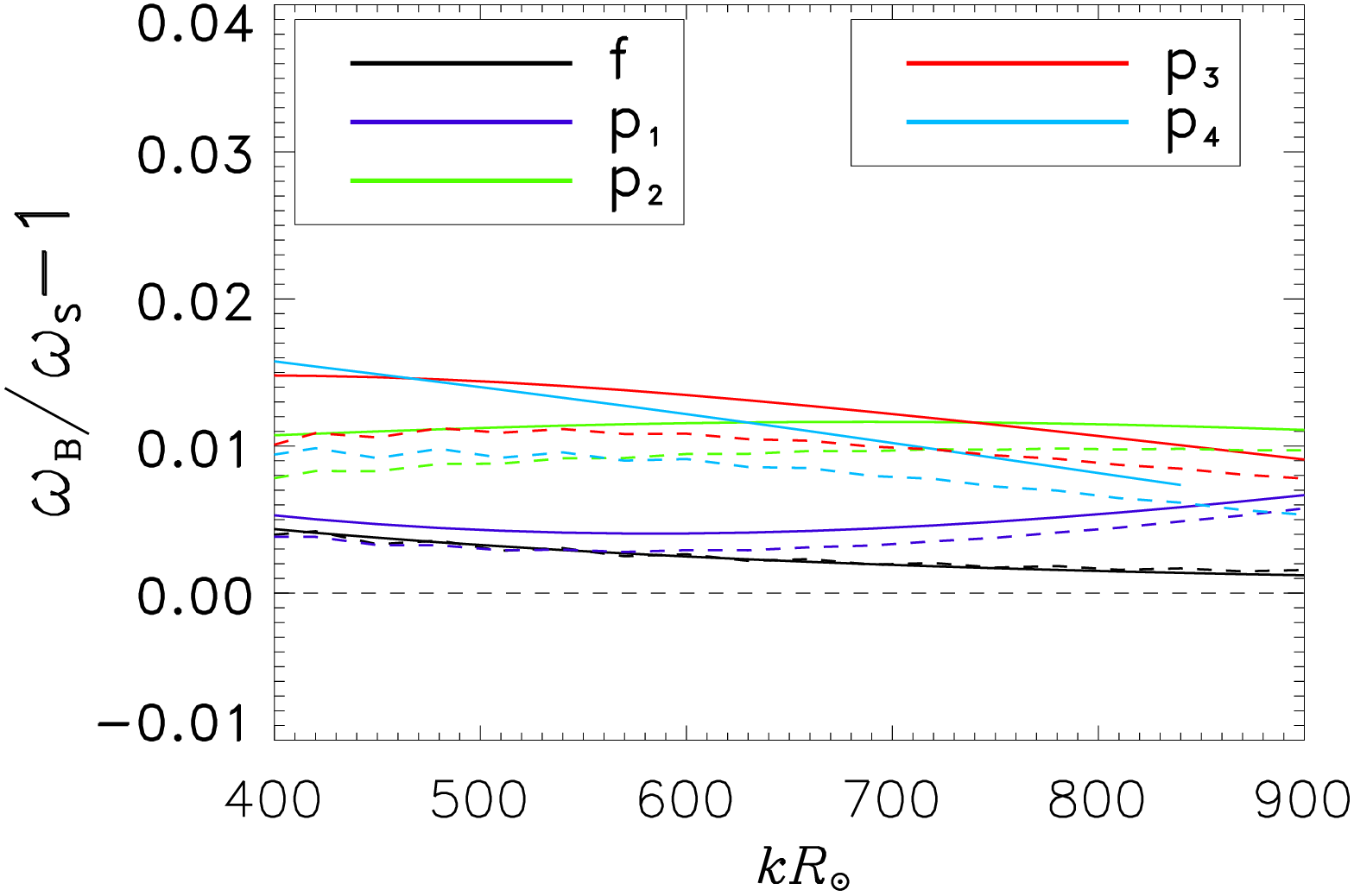}
\hspace{0.1cm}
\includegraphics[width=0.47\textwidth]{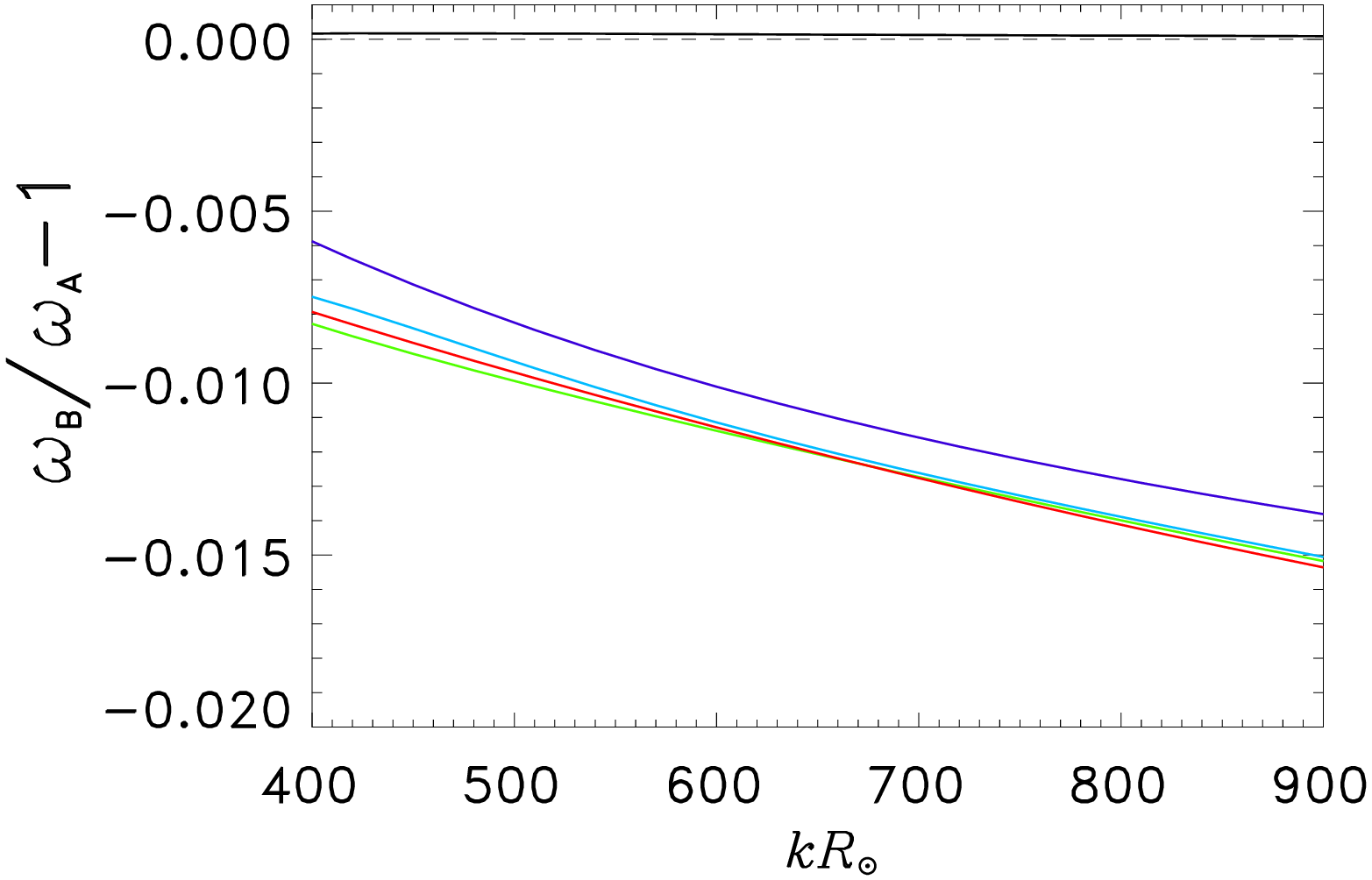}
\caption{Left: the relative difference between the real part of the CSM\_B eigenfrequencies, $\omega_B$, and the real part of the Model~S eigenfrequencies, $\omega_\mathrm{S}$   as a function of $kR_\odot$. The solid curves are differences in the eigenfrequencies calculated using \textsf{SLiM} simulations and the dashed curves are from the \textsf{BVP}.  Right: the relative frequency difference of the real part of the eigenfrequencies,   \rev{calculated by \textsf{SLiM}} as a function of wavenumber  between CSM\_B and CSM\_A brought about by the reduction in sound speed.}
\label{efreqb}
\end{center}
\end{figure}

The eigenfunctions  are slightly adversely affected as can be seen by comparing Figure~\ref{smsb_efunc} with Figure~\ref{smsa_efunc}, however they are still more solar-like than those of CSM (Figure~\ref{nocs_efunc}). 
The  real parts of the eigenfrequencies (Figure~\ref{efreqb}) reduce to within 1\% of  Model~S. 
We have not found a model which resulted in \rev{more similar} eigenfrequencies without grossly changing the eigenfunctions.
With this sound speed profile, we have arrived at a convectively stable model, CSM\_B, with  eigenfrequencies  \rev{closer to those of Model~S} than CSM or CSM\_A. 


\section{Comparison of Eigenfunctions}\label{sl}

\begin{figure}
\begin{center}
\includegraphics[width=0.9\textwidth]{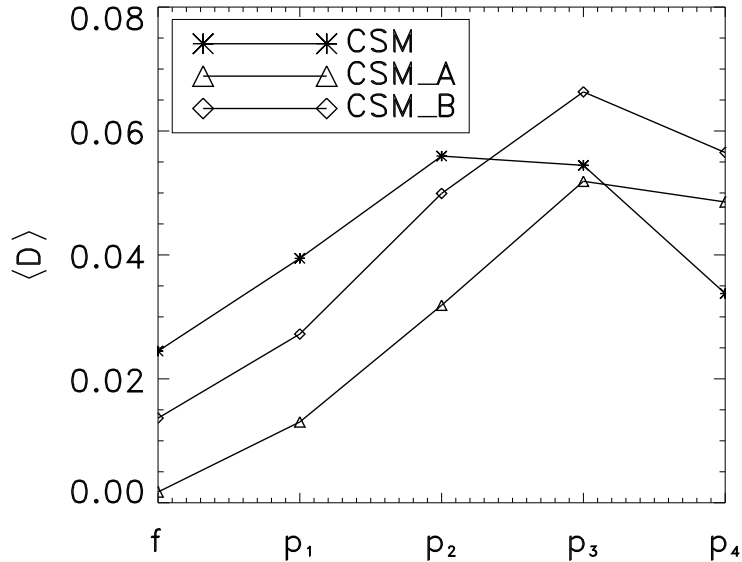}
\caption{The relative difference in area under $\textrm{Re}[v_z\sqrt{\rho}]$ (see Figures~\ref{nocs_efunc},\ref{smsa_efunc},\ref{smsb_efunc})   averaged over $400 \le kR_\odot \le 900$, $\langle D\rangle$, between each background model - CSM (asterisk), CSM\_A (diamond) and CSM\_B (triangle) - and  Model~S  as a function of radial order.}
\vspace{-1cm}
\label{avec}
\end{center}
\end{figure}

Quantitatively, we compare the eigenfunctions by finding the relative difference of the area under $\mathrm{Re}[v_z\sqrt{\rho}]$ between  Model~S and each convectively stable background in the near-surface layers, $-1.0~\mathrm{Mm} \le z \le 0.5$~Mm. The difference is defined by
\begin{equation}
D = \frac{   \int_{-1~\mathrm{Mm}}^{0.5~\mathrm{Mm}} (    \mathrm{Re}[v_z \sqrt{\rho}]  \, - \mathrm{Re}[v_{z\mathrm{S}} \sqrt{\rho_\mathrm{S}}] )^2 \, \textrm{d}z }  {  \int_{-1~\mathrm{Mm}}^{0.5~\mathrm{Mm}} (  \mathrm{Re}[v_{z\mathrm{S}} \sqrt{\rho_\mathrm{S}}])^2 \, \textrm{d}z}.
\label{unorm}
\end{equation}
This integration range was chosen because this is where the stabilisation has greatest effect. For each radial order we take the mean of $D$ over $400 \le kR_\odot \le 900$, giving  a quantitative measure \rev{of the differences between the eigenfunctions of Model~S and the stabilised model} [$\langle D \rangle$].  Figure~\ref{avec} shows that for the \textit{f}, \textit{p}$_1$  and \textit{p}$_2$ modes CSM\_A (triangle) has eigenfunctions closest to that of Model~S, while CSM (asterisk) has those farthest from Model~S.

\begin{figure}
\begin{center}
\includegraphics[width=0.9\textwidth]{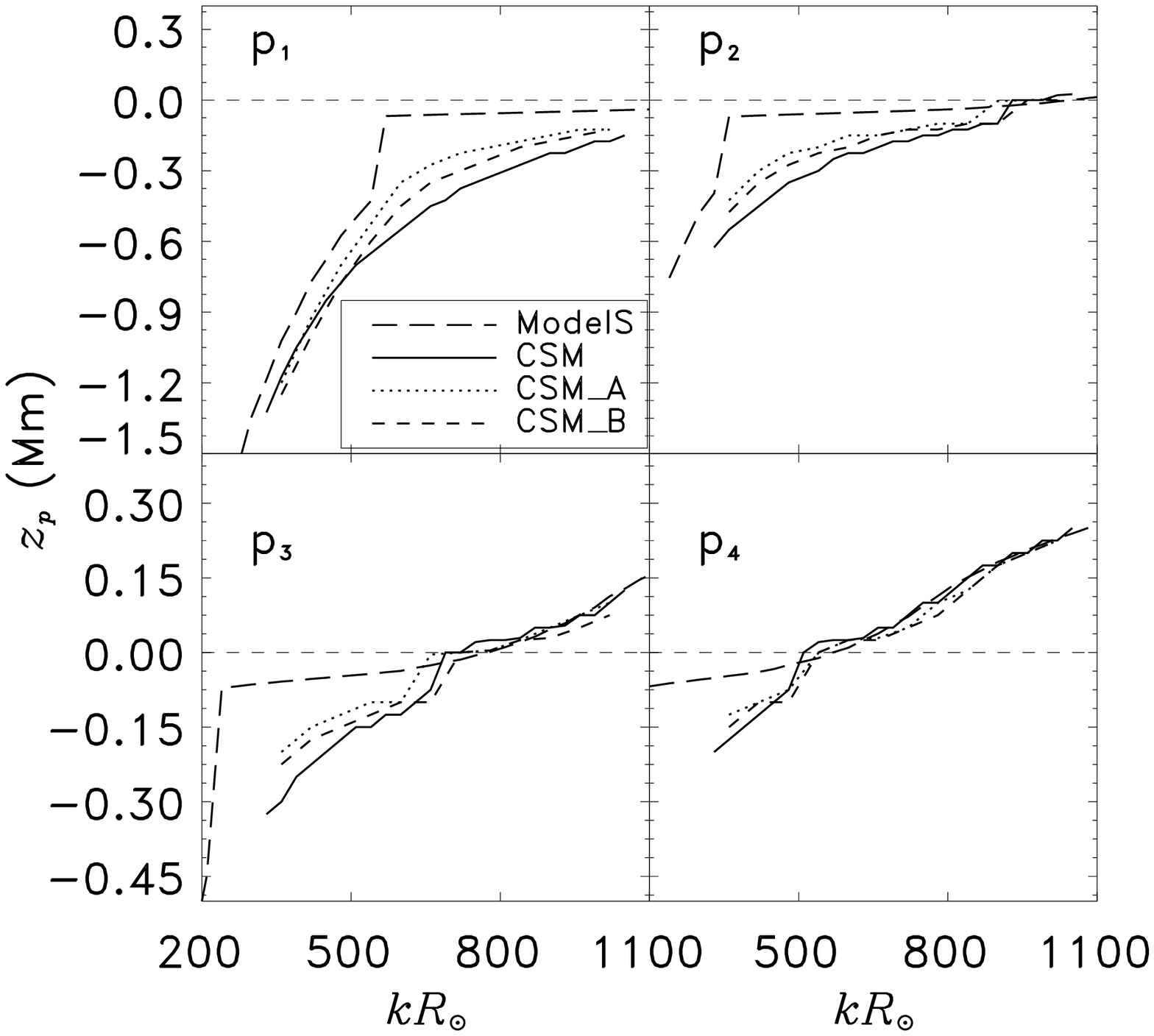}
\caption{The height of the  uppermost peak of $\textrm{Re}[v_z\sqrt{\rho}]$ (see Figures~\ref{nocs_efunc},\ref{smsa_efunc},\ref{smsb_efunc}) for each radial order as a  function of $kR_\odot$. The stabilisation reduces the height of the peak from Model~S (long-dash) to CSM (solid). The consequence of adjusting the sound speed is shown in CSM\_A (dot) and CSM\_B (short-dash).}
\label{pken}
\end{center}
\end{figure}

In addition,  we measure the height of the uppermost peak, $z_p$, of $\textrm{Re}[v_z\sqrt{\rho}]$.
Figure~\ref{pken} shows $z_p$ for each radial order and each model as indicated. From this we see that stabilising the background causes $z_p$ to drop in height (\textit{i.e.} the difference between the solid and the long-dash curves). Increasing the sound speed in a narrow region close to the surface (CSM\_A) pushes the peak back towards the surface (dotted curves). 
The broad decrease in sound speed added in CSM\_B does not change the location of the peak too much (short-dash curves). The sudden transition to very high upper turning points  at high wavenumber for Model~S (particularly for the \textit{p}$_1$ and \textit{p}$_2$ modes)  is due to the protuberance in the Model~S eigenfunctions very close to the surface (for example, the  \textit{f} and \textit{p}$_1$-modes in Figure~\ref{nocs_efunc}) which is absent in the stable models. 
The protuberance is due to rapid changes in the density scale height close to the surface that disappears after the stabilisation (reduction of $\mathrm{d}_zp$).

We now have three convectively stable solar models \rev{each having similar, but slightly different, eigenfrequencies and eigenfunctions to Model~S.}
Having models focused on achieving slight variations of the same goal (\rev{more similar} eigenfunctions or eigenfrequencies) gives us the possibility of testing the sensitivity of helioseismic analysis techniques to the background properties.


\section{Modelling the Random Wave Field}

\subsection{Random wave excitation model}\label{rwem}
We model the random wave excitation by imposing a vertical force, $f_z$, to the right-hand-side of Equation~(\ref{eqn:F}). The force is specified by
\begin{equation}
f_z(\mathbf{k}_i,z,\omega_j)= \rho G_{ij} e^{-(z-z_{*})^2/d^2},
\end{equation}
where $\mathbf{k}_i$ is a horizontal wavevector, $\omega_j$ is an angular frequency,  $d=0.075$~Mm is the width of the source, and the acceleration $G_{ij}$ is a realisation of a complex Gaussian random variable with zero mean and variance  $E[|G_{ij}|^2] = \mathrm{exp}\left[  -(\omega_j)^2/2\sigma^2 \right]$ where $\sigma/2\pi=2.12$~mHz \citep{Gizon2004}.  The height of the  sources is at $z_{*}=- 0.75$~Mm, which is close to the highly superadiabatic layer where  solar waves are expected to be strongly excited \citep{Nigam1999}.  
In reality, the sources in the Sun will also have a wavenumber dependence which we have not included. In practice, the sources are generated before the simulation commences and saved with a 30~second cadence. 
The forcing is applied at each time step (in cases herein this is approximately 0.13  solar seconds), with the value of the applied forcing changing every 30 solar seconds.  
We remark that we first tried to use a Lorentzian for  the  frequency dependence \citep{Title1989,Gizon2002}, corresponding to sources which decay exponentially in time. 
We found that the resulting power was too strong  at high frequencies compared with observations, and that the Gaussian distribution produced a  better agreement. 

\subsection{Azimuthally Averaged Power Spectra}

In this section we used \textsf{SLiM} to investigate the response of CSM\_A and CSM\_B to the random wave excitation model as described in Section~\ref{rwem}. 
A total of 16 hours was simulated, however the  first eight hours, during which the wave field is reaching a steady state, are discarded. To mimic SOHO/MDI observations, we save vertical-velocity data at a height of 0.2~Mm above the surface (the height at which SOHO/MDI observes, see  \cite{Bruls1993}) and account for the modulation transfer function of the instrument by multiplying the simulated power spectra by the modulation transfer  function of \cite{RabelloSoares2001}.

To make a comparison with an observed power spectra, we took eight hours of Postel projected (centred at a longitude of $170^\circ$ and latitude of $-8.3^\circ$)  full-disk Doppler observations with a 60~second cadence from SOHO/MDI on 21 January 2002. 
The observations consist primarily of quiet Sun covering a surface area  identical to  the simulations. 

\begin{figure}
\begin{center}
\includegraphics[width=1.0\textwidth]{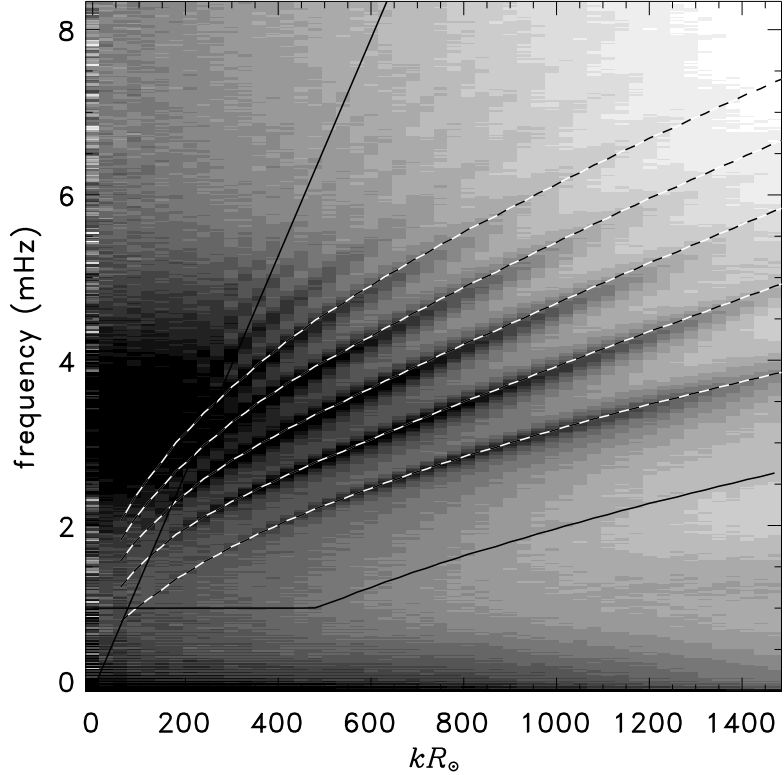}
\caption{The azimuthally averaged power spectrum of  eight hours of quiet-Sun SOHO/MDI Doppler observations. The eigenfrequencies of Model~S are the dashed curves. The straight solid line and the bottom solid curve form the boundaries of the comparable domain  of the simulations. Stronger power is black and weaker power is white.}
\label{OBSSPECTRA}
\end{center}
\end{figure}

\begin{figure}
\begin{center}
\includegraphics[width=1.0\textwidth]{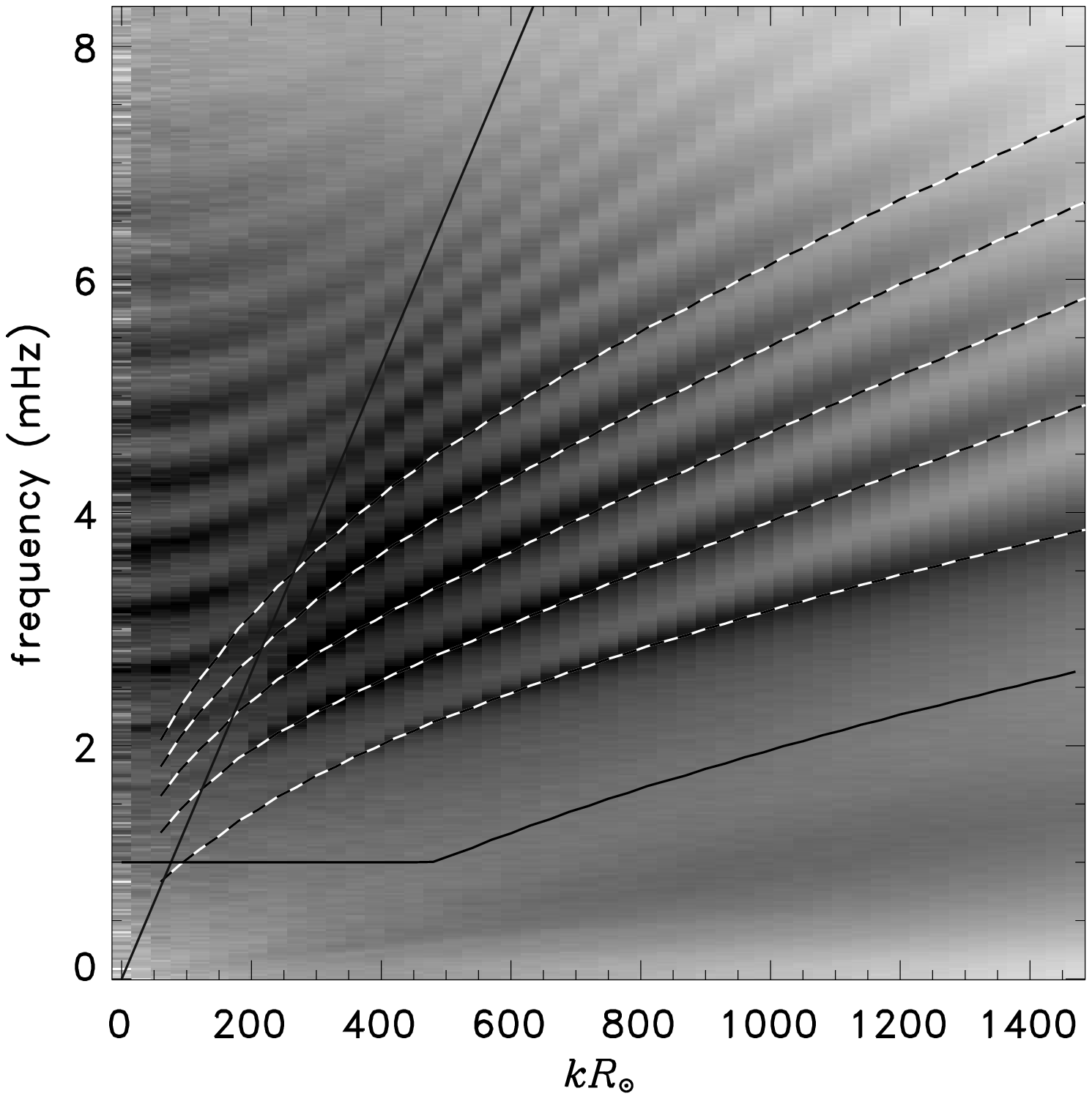}
\caption{The azimuthally averaged power spectrum of  eight hours of simulated random wave excitation in CSM\_A, accounting for the SOHO/MDI modulation transfer function and presented on the same log-power scale as Figure~\ref{OBSSPECTRA}. The eigenfrequencies of Model~S are the dashed curves. The straight solid line is where  $\omega/k$ is equal to $c_A(z_b) /(1 + z_b/R_\odot)$, and the bottom solid curve [$b(k)$] form the boundaries of the comparable domain. }
\label{ASPECTRA}
\end{center}
\end{figure}

\begin{figure}
\begin{center}
\includegraphics[width=1.0\textwidth]{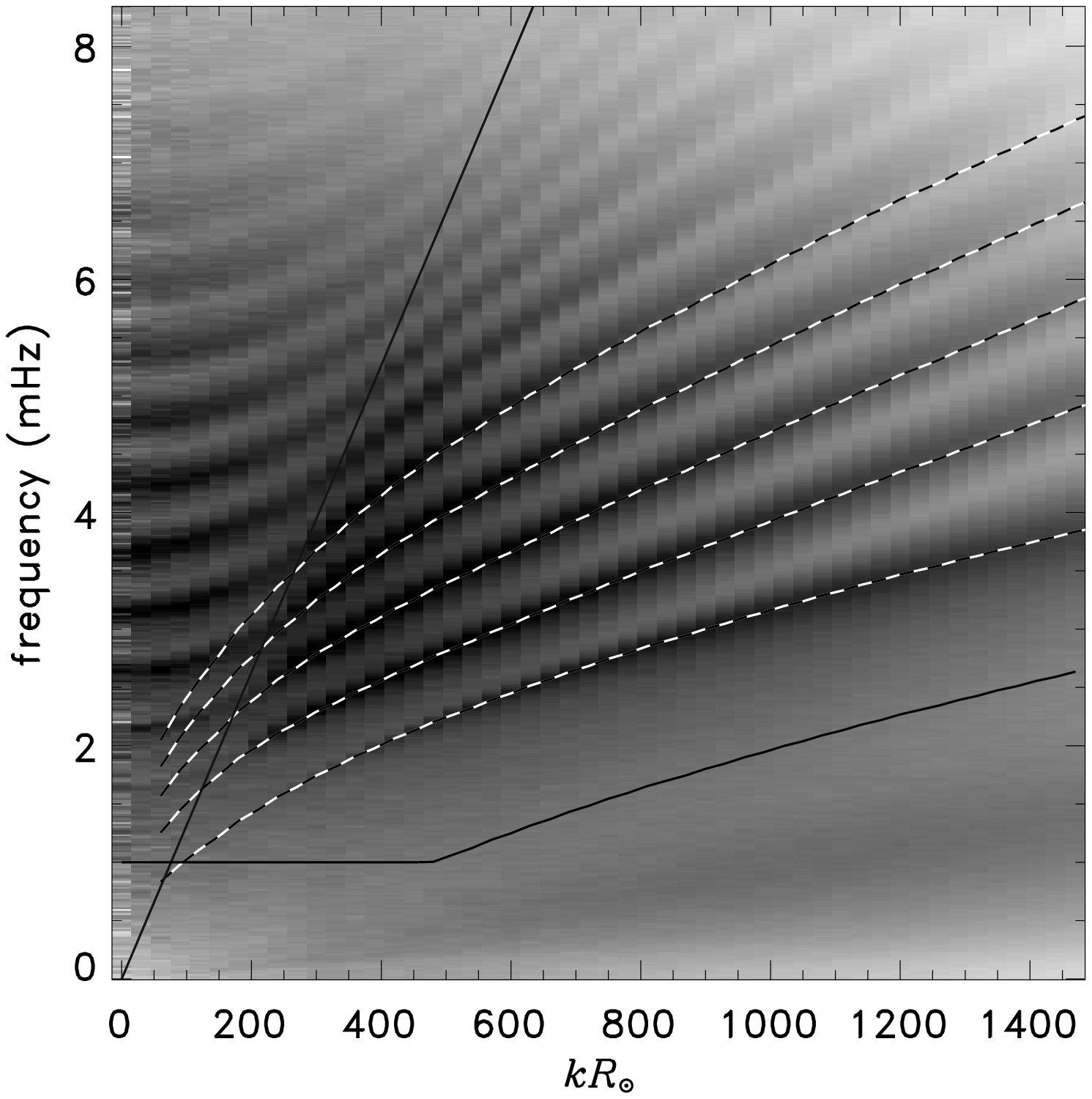}
\caption{The azimuthally averaged power spectrum of  eight hours of simulated random wave excitation in CSM\_B, accounting for the SOHO/MDI modulation transfer function  and presented on the same log-power scale as Figures~\ref{OBSSPECTRA} and \ref{ASPECTRA}. The eigenfrequencies of Model~S are the dashed curves. The straight solid line is where  $\omega/k$ is equal to $c_B(z_b) /(1 + z_b/R_\odot)$, and the bottom solid curve [$b(k)$] form the boundaries of the comparable domain. }
\label{BSPECTRA}
\end{center}
\end{figure}

We consider the azimuthally averaged (with bin size $\Delta k =2 \pi/[145.77$~Mm]) power spectra of the observations, $P(k_x,k_y,\omega)=| v_\mathrm{los}(k_x,k_y,\omega) |^2$, and of the simulations with CSM\_A and CSM\_B, $P(k_x,k_y,\omega)=| v_{z}(k_x,k_y,\omega) |^2$ are shown in Figures~\ref{OBSSPECTRA}, \ref{ASPECTRA}, and \ref{BSPECTRA}  respectively.
The dashed curves are the eigenfrequencies calculated from Model~S \rev{for comparison.} 
The straight solid line is where $\omega/k$ is equal to $c(z_b) /(1 + z_b/R_\odot)$ and $z_b=-22.6$~Mm; as stated previously, modelling a higher $\omega/k$ would require a deeper box.
There is some power evident in the low frequencies
which are most likely \textit{g}-modes introduced by stabilising the background.
\rev{
These are the artificial product of having a stable model.
Thus, this region cannot be compared to solar observations.
The remaining
}
``comparable domain'': $ b(k) < \omega/2\pi < k \, c(z_b) /(1 + z_b/R_\odot)$ where $b(k)$ is the lower curve shown in these figures, 
\rev{should contain modes which are comparable to those on the Sun.}
The  azimuthally averaged power spectra are normalised to the mean power within a region defined by $(kR_\odot-600)^2/200^2 + (\omega/2\pi - 3~\textrm{mHz})^2/(1~\textrm{mHz})^2 \le 1$. 
By inspection, the power spectra of CSM\_A (Figure~\ref{ASPECTRA}) and CSM\_B (Figure~\ref{BSPECTRA}) look qualitatively similar to the observed spectrum (Figure~\ref{OBSSPECTRA}). 
We now take a closer look at the properties.

\subsection{Amplitudes of the Power Spectra}

\begin{figure}
\begin{center}
\includegraphics[width=0.7\textwidth,angle=90]{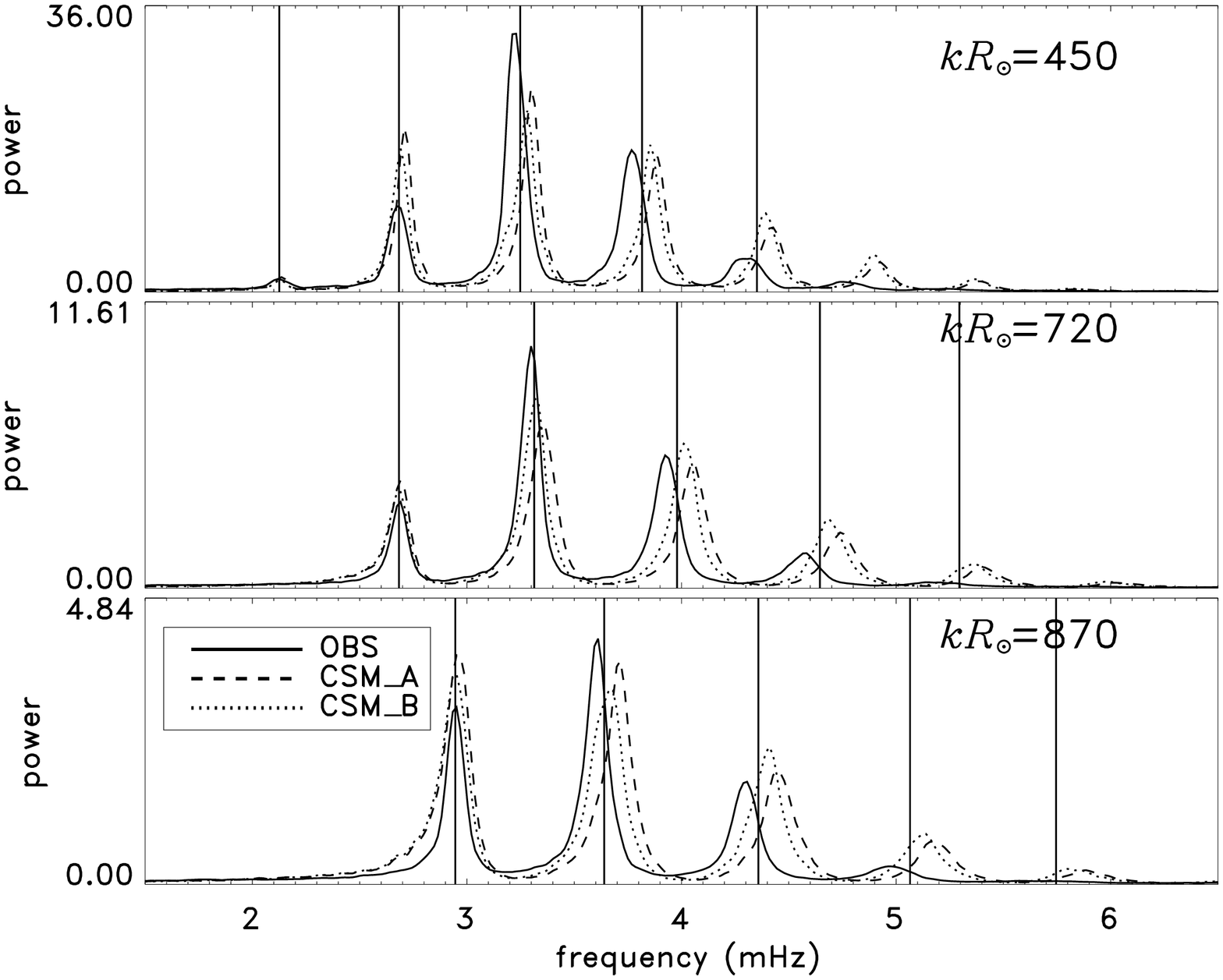}
\caption{Cuts (smoothed over 0.035~mHz for the purpose of this plot) through the azimuthally averaged power spectra in arbitrary units at the indicated wavenumbers for CSM\_A (dash), CSM\_B (dot) and observations (solid) as a function of frequency. The vertical black lines are the eigenfrequencies of  Model~S. }
\label{pcut}
\end{center}
\end{figure}

Figure~\ref{pcut} shows vertical cuts through the power spectra in Figures~\ref{OBSSPECTRA}, \ref{ASPECTRA}, and ~\ref{BSPECTRA} as a function of frequency. The Model~S eigenfrequencies (vertical lines) are larger than those of the observations (solid curve), while CSM\_A (dash curve) and CSM\_B (dot curve) eigenfrequencies are larger than those of Model~S. 
It also shows that the maximum power and linewidths of the ridges agree with observations best at low frequency.   

\begin{figure}
\begin{center}
\includegraphics[width=1.0\textwidth]{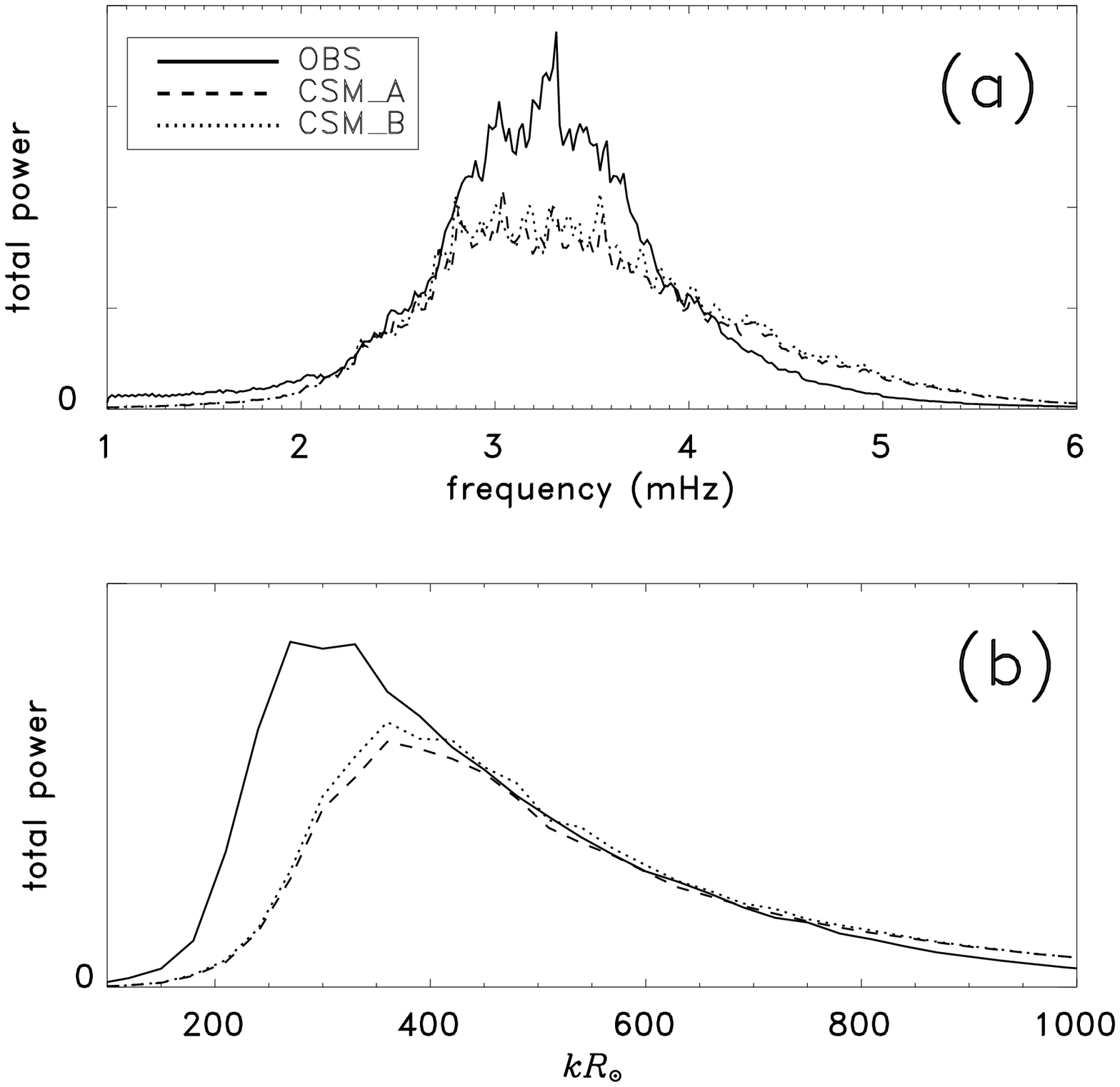}
\caption{The total of the azimuthally averaged power in the comparable range  as a function of  (a) frequency \revtwo{(averaged over  wavenumber in the comparable range)}  and (b) $kR_\odot$ \revtwo{(averaged over all frequency in the comparable range)}  for CSM\_A (dash), CSM\_B (dot) and observations (solid), in arbitrary units.}
\label{pave}
\end{center}
\end{figure}

Figure~\ref{pave} shows the total power in the comparable range for the observations (solid curve), CSM\_A (dash curve) and CSM\_B (dot curve) as a function of  (a) frequency and (b) $kR_\odot$.
The  maximum power in the simulations occurs at a larger wavenumber than in the observational power. 
Correcting this could be done by fine tuning the wave excitation model, and may be done in the future, however the results presented here are sufficiently close for a large number of studies.

\subsection{Fitting the Power Spectra}
We analyse the properties of the azimuthally averaged  power spectra in Figures~\ref{OBSSPECTRA}, \ref{ASPECTRA} and \ref{BSPECTRA}  by fitting asymmetric Lorentzians \cite[\textit{e.g.}][]{Duvall1993,Gizon2006},
\begin{eqnarray}
L ( \omega) = \sum_{n=0}^4 P_n \left[ \frac{( 1 + B_n X_n )^2 + B_n^2 }{ 1 + X_n^2 }  \right] + N
\label{alor}\\
\textrm{    where   }
X_n = \frac{ \omega - \omega_n}{\Gamma_n/2}
\textrm{      and      }
B_n = \frac{\Gamma_n/2}{\omega_{n} - \omega_\mathrm{v} }  \nonumber
\end{eqnarray}
 to cuts at fixed wavenumber as a function of frequency.
In Equation~(\ref{alor}), the maximum power of the $n^\textrm{th}$ ridge is given by $P_n$ and is located at a frequency $\omega_n$,  the valley is at $\omega_\mathrm{v}$, the noise is $N$, and the full-width-at-half-maximum (FWHM) of the asymmetric Lorentzian  is $\Gamma_n$. 
The fitting is done using a Levenberg-Marquardt algorithm for least squares curve fitting using the IDL \texttt{mpfit} package. 
The frequency range of the fit is from $\approx 0.6$ of the \textit{f}-mode Model~S eigenfrequency to $\approx 1.1$ of the \textit{p}$_4$-mode   Model~S eigenfrequency. 
We define the asymmetry parameter as $\chi_n = B_n \omega_n / (\Gamma_n /2)$ \citep{Gizon2006}. 

\begin{figure}
 \includegraphics[width=0.9\textwidth]{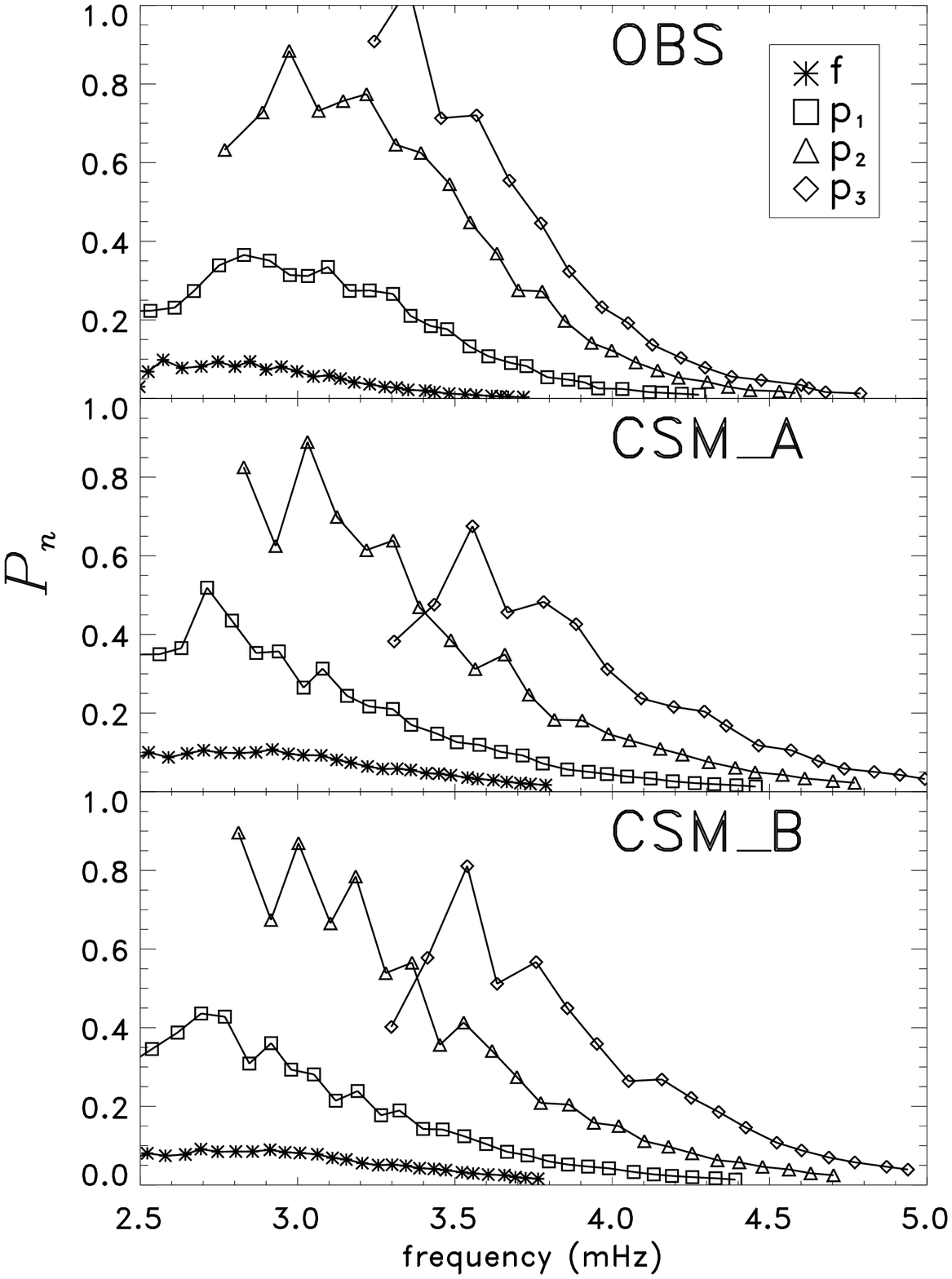}
\caption{The maximum power, $P_n$ for $n=0,1,2,3$ ridges calculated by fitting Equation~(\ref{alor}) to the azimuthally averaged power spectra as a function of frequency. The top panel shows results from the observations, the middle panel from CSM\_A and the bottom panel from CSM\_B. Each ridge is presented by a different symbol as indicated in the legend.}
\label{power}
\end{figure}

Figure~\ref{power} shows the maximum power of each $n$ from fitting Equation~(\ref{alor}) to the  power spectrum of the observations (top), CSM\_A (middle) and CSM\_B (bottom). The simulated power spectra have stronger power at high frequency than the observations.
In addition, the maximum power of $n=1$ occurs at a lower frequency in the simulations than in the observations.

\begin{figure}
 \includegraphics[width=0.9\textwidth]{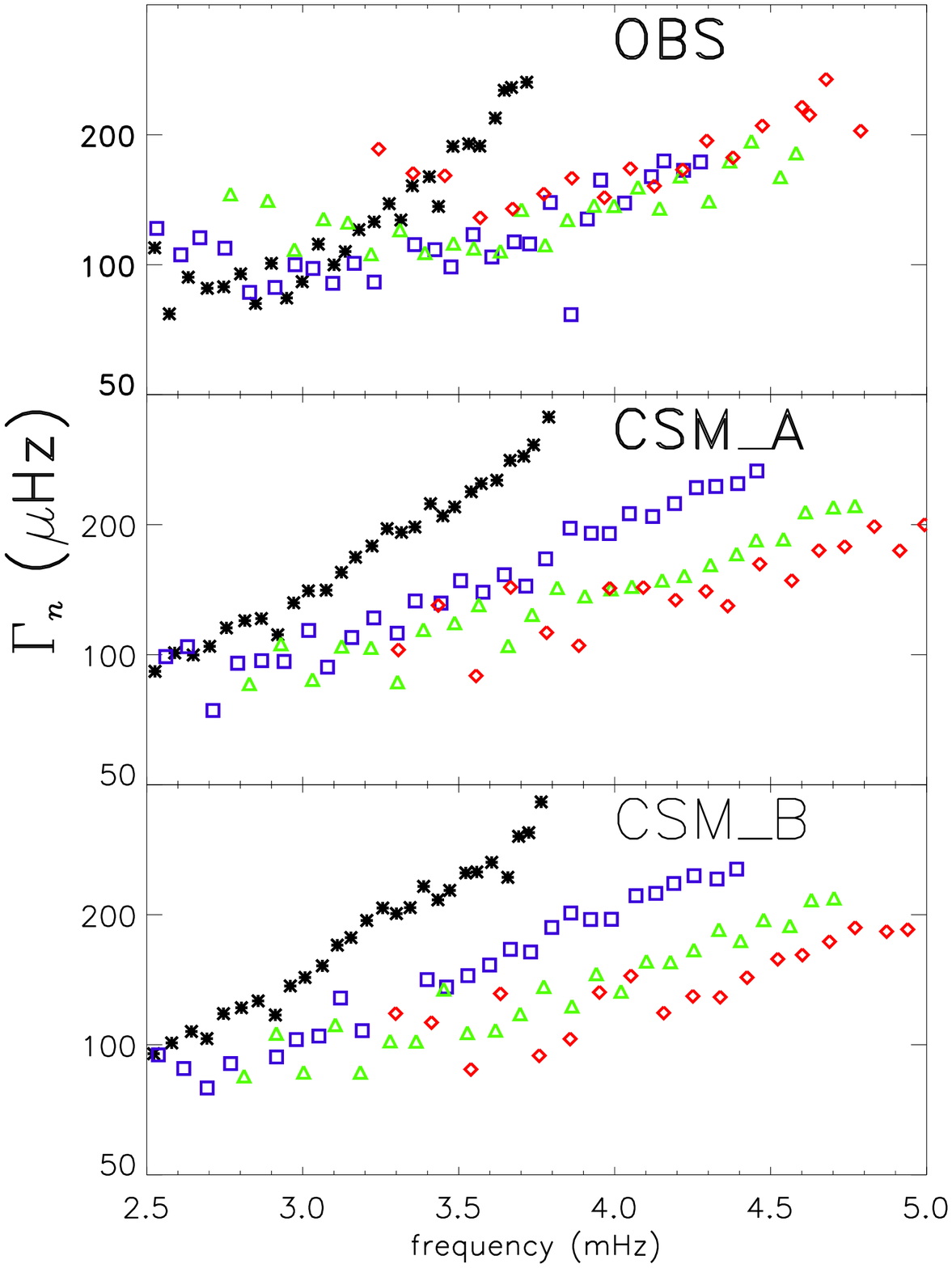}
\caption{The FWHM, $\Gamma_n$, for $n=0,1,2,3$ ridges  calculated by fitting Equation~(\ref{alor}) to the azimuthally averaged power spectra as a function of frequency. The top panel shows results from the
observations, the middle panel from CSM\_A and the bottom panel from CSM\_B. The symbol legend is the same as in Figure~\ref{power}.}
\label{fwhm}
\end{figure}
Figure~\ref{fwhm} shows the FWHM of the Lorentzian fit for each mode in the  power spectrum of the observations (top), CSM\_A (middle) and CSM\_B (bottom). The FWHM of the ridges in the observations is consistent with Figure~2 in \cite{Antia1999}, keeping in mind that these are coarse measurements. The simulation ridges have larger FWHMs than the observations for \textit{f} and \textit{p}$_1$ modes.

\begin{figure}
\begin{center}
\includegraphics[width=0.9\textwidth]{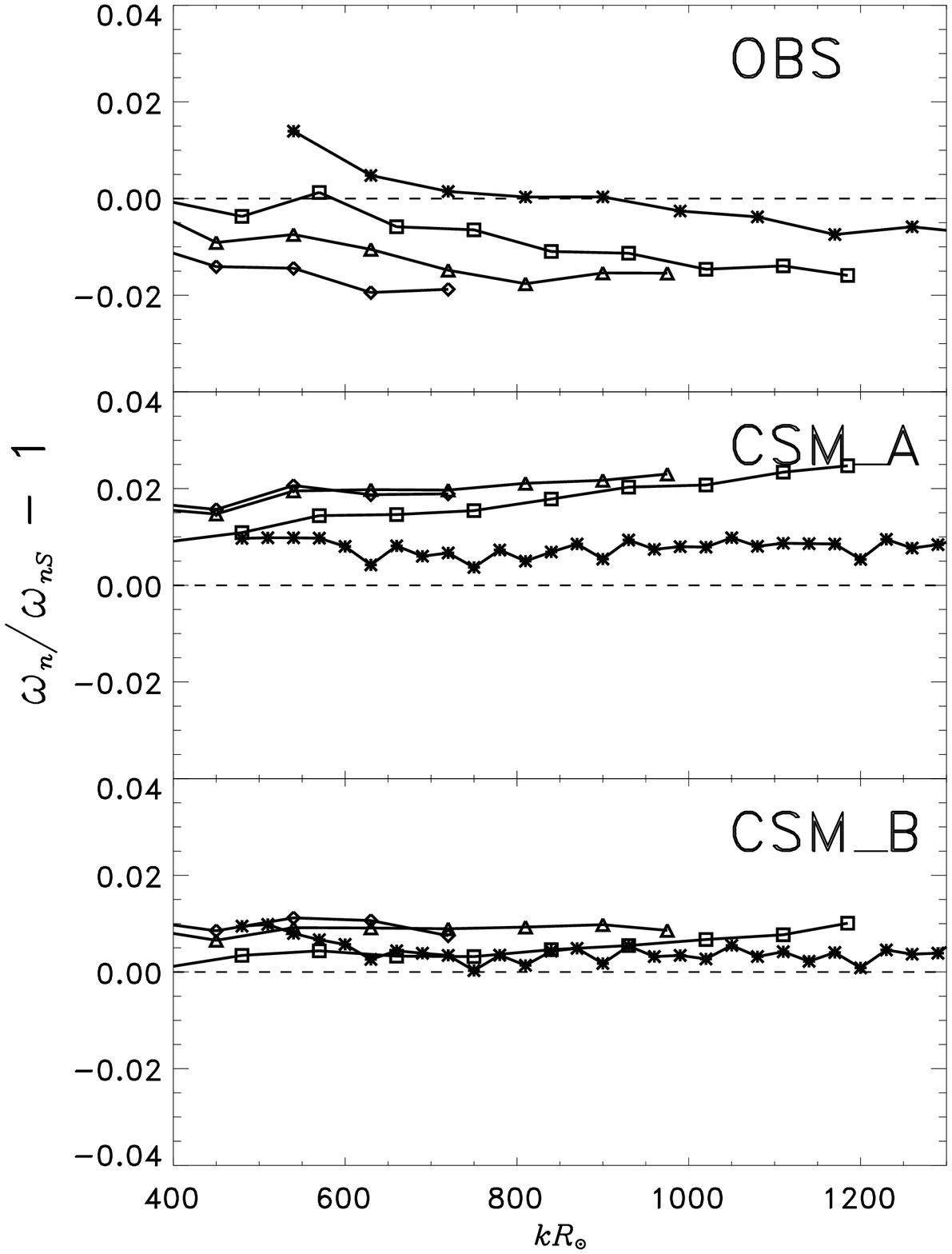}
\caption{The relative difference of the central ridge frequencies [$\omega_n$] for $n=0,1,2,3$ ridges  calculated by fitting Equation~(\ref{alor}) to the azimuthally averaged power spectra,  to those of Model~S as a function of $kR_\odot$.  The symbol legend is the same as in Figure~\ref{power}. The top panel shows results from the observations, the middle panel from CSM\_A, and the bottom panel from CSM\_B.}
\label{freq}
\end{center}
\end{figure}
\begin{figure}
\begin{center}
\includegraphics[width=0.9\textwidth]{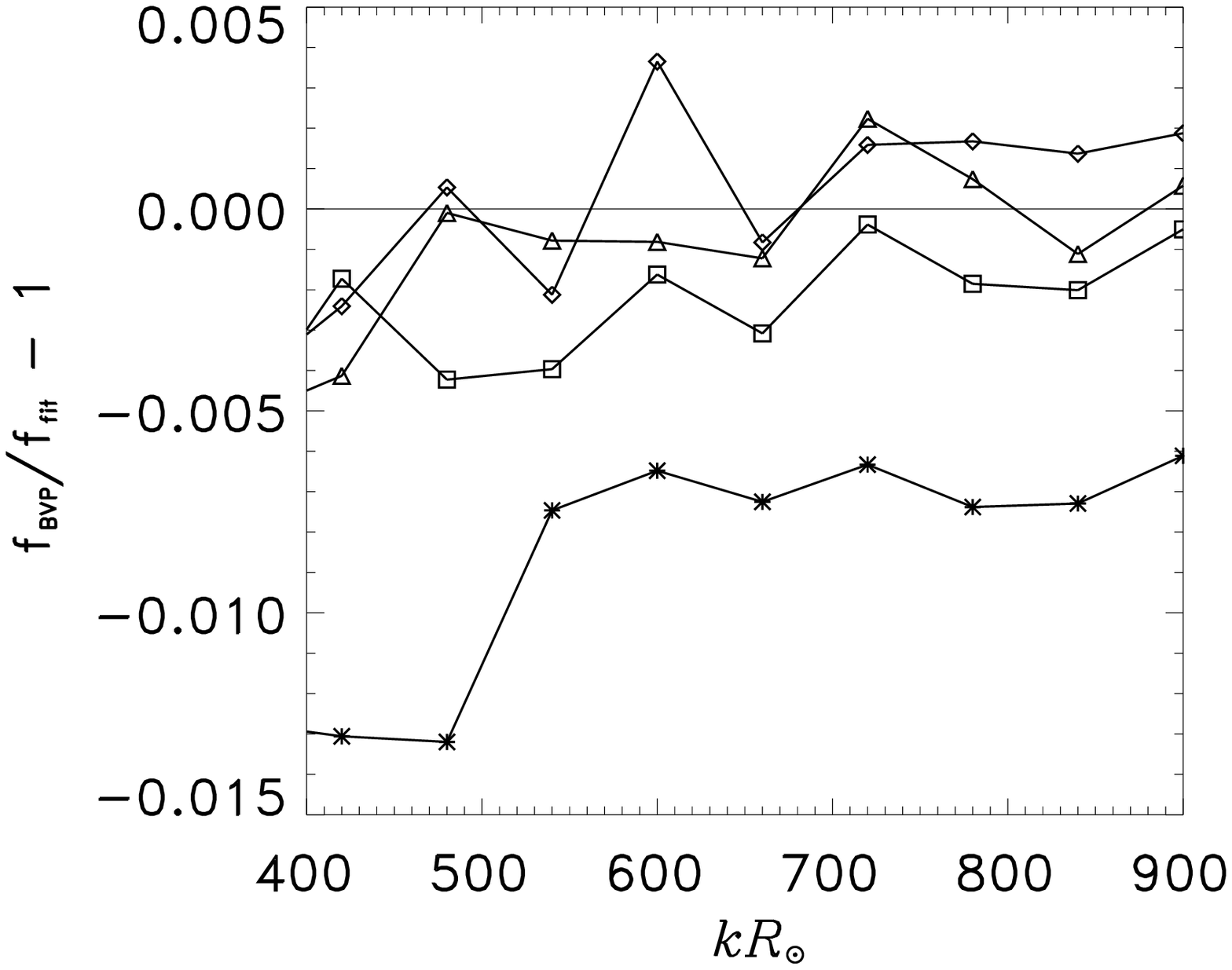}
\caption{
\rev{The relative difference between  the eigenfrequencies of the \textsf{BVP} solutions, $f_\mathrm{BVP}$, and the frequency of the maximum ridge power as identified from fitting the power spectrum, $f_\mathrm{fit}$, for CSM\_A.  The symbol legend is the same as in Figure~\ref{power}. }
}
\label{freqcomp}
\end{center}
\end{figure}

Figure~\ref{freq} shows the relative difference of the central ridge frequencies to Model~S for the observations (top), CSM\_A (middle) and CSM\_B (bottom).  \rev{The results from the Lorentzian fitting are within 1\% of the \textsf{BVP} solutions as shown in Figure~\ref{freqcomp}}. 

\begin{figure}
\begin{center}
\includegraphics[width=0.9\textwidth]{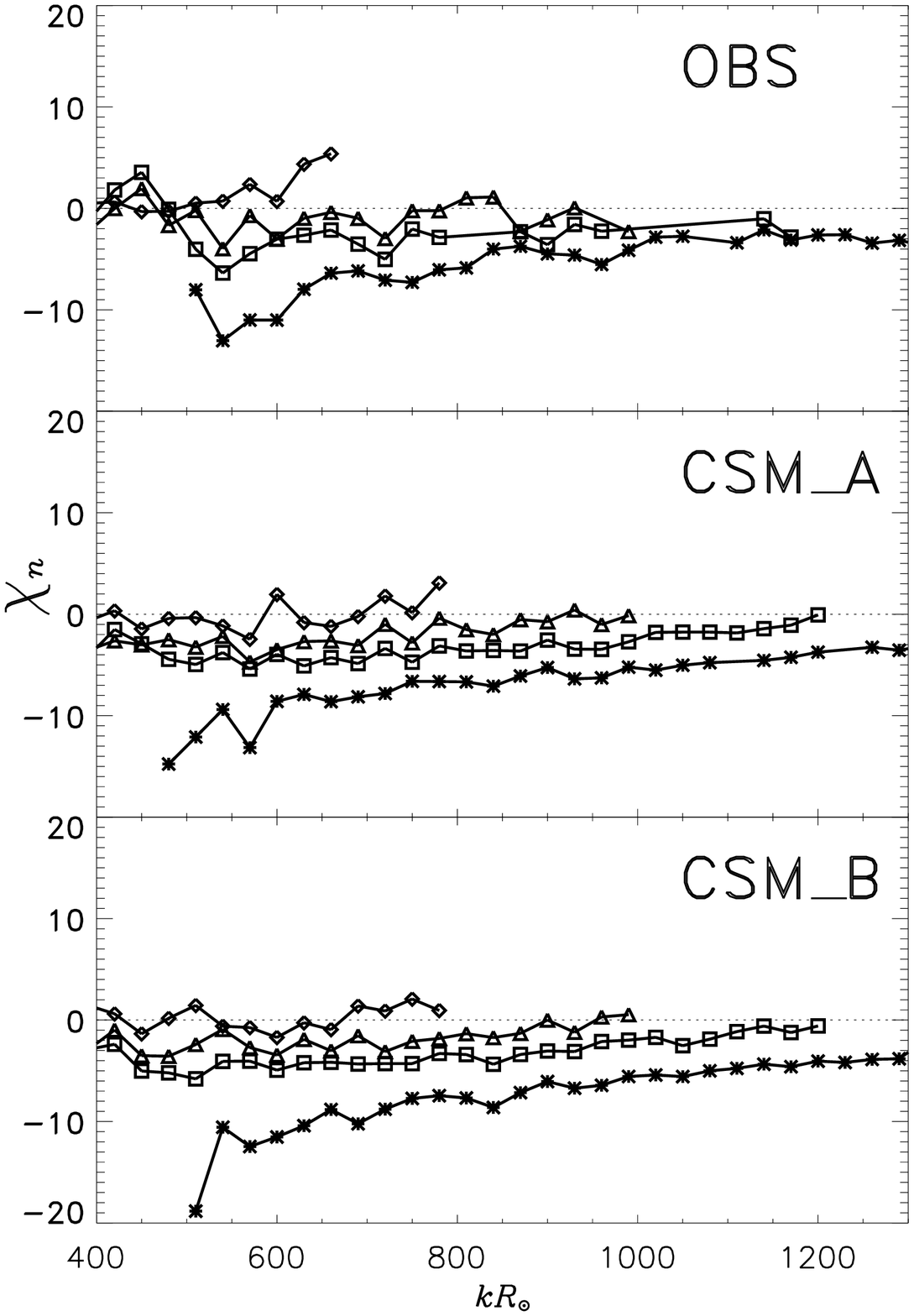}
\caption{The $\chi_n$ asymmetries for $n=0,1,2,3$ ridges calculated by fitting Equation~(\ref{alor}) to the azimuthally averaged power spectra  as a function of $kR_\odot$.  The symbol legend is the same as in Figure~\ref{power}. The top panel shows results from the observations, the middle panel from CSM\_A and the bottom panel from CSM\_B.}
\label{chi}
\end{center}
\end{figure}

Figure~\ref{chi} shows the $\chi_n$ asymmetries of the observations (top), CSM\_A (middle) and CSM\_B (bottom). We achieve the correct sign and comparable magnitude of the asymmetry for all the modes. The \textit{f}-mode has negative asymmetries, and the value of the asymmetries increases with increasing mode number which is in agreement with \cite{Gizon2006}.

We have demonstrated the \rev{response}  of the numerical simulations of wave excitation in the Sun using two of the convectively stable background models, CSM\_A and CSM\_B. The eigenmodes of the background models and the parameters of the sources of acoustic wave oscillations are sufficient to be used as a foundation for quantitative solar-like simulations.

\rev{In addition, we have successfully implemented the stable background models into the framework of another code which also computes linear simulations of helioseismic wave propagation, the Seismic Propagation through Active Regions and Convection (SPARC)  code \citep{Hanasoge2006,Hanasoge2007}.
}

\section{Discussion}
We have created three convectively stable solar models which, to slightly differing extents,  have similar  eigenmodes  to those of Model~S. We have also computed helioseismic simulations using a model for the random excitation of
waves, which together with the stable solar models, reproduce the SOHO/MDI observed  mode frequencies and  asymmetries well for each of the \textit{f} and \textit{p}$_1$ to \textit{p}$_4$ ridges. 
The linewidths of the ridges and the power distribution  are reasonably similar to those of the Sun. 

Although stabilising the background model is an important step in numerical studies of wave propagation \citep[and has been done before, \textit{e.g.} by][]{Parchevsky2007,Cameron2008,Shelyag2008,Schunker2010}, its effects on the eigenfunctions and eigenfrequencies has received little attention.  
An optimal way to produce a  convectively stable background model for numerical simulations has not been formulated,  but nevertheless the models presented here should be useful for a range of studies. 
In particular, we envisage that these models will be used to study the propagation of solar waves through three-dimensional heterogeneities, such as convective flows, granulation and model sunspots \citep[\textit{e.g.}][]{Cameron2010,Dombroski2011}. Having three models with slightly different properties will enable us to quantitatively test the sensitivity of the results to the details of the models. 
The models \rev{and extra information from the analysis in this paper} are available for download from the HELAS local helioseismology website \citep[\url{http://www.mps.mpg.de/projects/seismo/NA4/};][]{Schunker2008}.

\clearpage

\appendix


\section{Calculating Eigenmodes Using Simulations} \label{app1}
The following procedure calculates the eigenmodes of the system using \textsf{SLiM} and is designed  to be applied iteratively.  
We  began by simulating the response of  the system to a wave packet constructed from Model~S eigenmodes of one radial order (as in Cameron, Gizon, and Duvall, 2008).  
The outputs, $v_x(k,z,t)$ and $v_z(k,z,t)$, of a five hour long simulation were saved with a one minute cadence.  
We then took the Fourier transform of the velocity field in time, $v_z(k,z,\omega)$.
From this we determined the eigenfrequencies  from a linear fit in time to the phase $\phi(k,t)=\textrm{Arg}[{v_z}(k,z=200~\mathrm{km},t)]$ with the $2\pi$ wrap-around removed. 
The function we used to fit the phase, $\phi(k,t)$, is given by  $\phi(k,t)=\mathrm{Re}[\omega_\mathrm{i} (k)] t + \phi_\mathrm{off}(k)$.
The height of 200~km corresponds to the observation height of SOHO/MDI  \citep{Bruls1993}.
\rev{We determined the radial component $v_{z\mathrm{f}}(k,z,\omega)$ by applying a broad ridge filter to isolate the appropriate radial order, centred on the improved (subscript $\mathrm{i}$) estimate of the real part of the eigenfrequencies $\mathrm{Re}[\omega_\mathrm{i}(k)]$.}
The same filter was applied to the horizontal velocity to get  $v_{x\mathrm{f}}(k,z,\omega)$. 
\rev{The velocities are then Fourier transformed from frequency space back to time.}

The improved eigenmodes are then given by 
\begin{eqnarray}
v_{x\mathrm{i}}(k,z) = \frac{ \int_{0}^{300\mathrm{min}} v_{x\mathrm{f}}(k,z,t) \exp[-i (\omega_\mathrm{i}(k) t + \phi_\mathrm{off}) ] dt  }
                              { \int_{0}^{300\mathrm{min}} v_{z\mathrm{f}}(k,z=200~\mathrm{km},t) \exp[-i (\omega_\mathrm{i}(k) t + \phi_\mathrm{off}) ] dt   } \nonumber \\
v_{z\mathrm{i}}(k,z) = \frac{ \int_{0}^{300\mathrm{min}} v_{z\mathrm{f}}(k,z,t) \exp[-i (\omega_\mathrm{i}(k) t + \phi_\mathrm{off}) ] dt  }
                              { \int_{0}^{300\mathrm{min}} v_{z\mathrm{f}}(k,z=200~\mathrm{km},t) \exp[-i (\omega_\mathrm{i}(k) t + \phi_\mathrm{off}) ] dt   }.      \nonumber
\end{eqnarray}
Note that $\phi_\mathrm{off}(k)$ and the denominator are defined from the vertical velocity component,  $v_{z\mathrm{i}}(k,z)=1$  at $z=200$~km.
From these eigenmodes we constructed a new wave packet initial condition and the simulation was re-computed with this wave packet. 
In practice we found that  a single pass is sufficient and the improved eigenmodes from the first simulation were used to \rev{compare to Model~S}.

\section{Determining the Eigenmodes of the Boundary Value Problem}\label{app2}

The perturbations of a particular eigenmode with radial order $n$ of the Equation~(\ref{eqn:F}) have the form
\begin{eqnarray}
\mathbf{v}(\bk ,z,t) & =  &\left[ v_{z}(\bk ,z)\unitz  + v_{x}(\bk , z)\mathbf{\hat{x}} \right] e^{-i( \omega_t - \bk \cdot \ \mathbf{x})} \\
p^\prime (\bk , z,t) & =  & p^\prime (\bk , z)  e^{-i( \omega t - \bk \cdot \ \mathbf{x})} 
\end{eqnarray}
with $\mathbf{v}_n = (\partial_t + \gamma) \bxi_n $.  

After some manipulation (using the continuity equation, equation of motion and \revtwo{energy} equation), our system of equations becomes
\begin{eqnarray}
\rho \beta v_z = -\frac{\mathrm{d}p^\prime}{\mathrm{d}z} - g\left[  \frac{\rho v_z }{\beta^2}\frac{\mathrm{d} \gamma}{\mathrm{d}z}  - \frac{v_z}{\beta}\frac{\mathrm{d} \rho}{\mathrm{d}z} - \frac{\rho}{\beta}\frac{\mathrm{d} v_z}{\mathrm{d}z} -\frac{k^2 p^\prime}{ r^2\beta^2} \right] \label{f1}\\
p^\prime=-c^2\left[ \frac{2 \rho v_z}{r \beta} - \frac{\rho v_z}{\beta^2}\frac{\mathrm{d} \gamma}{\mathrm{d}z} + \frac{\rho}{\beta} \frac{\mathrm{d} v_z}{\mathrm{d}z} + \frac{p^\prime k^2}{r^2 \beta^2} \right] + \frac{v_z}{\beta}\frac{\mathrm{d} p}{\mathrm{d}z},
\label{f2}
\end{eqnarray}
where $\beta=\gamma - \mathrm{i}  \omega$.

Following the method of \cite{Birch2004}, we substitute
\begin{eqnarray*}
y_1=\frac{i p^\prime}{\sqrt{\rho c}}\\
y_2=v_z \sqrt{\rho c}
\end{eqnarray*}
into  Equations~(\ref{f1}) and (\ref{f2}) to get
\begin{eqnarray}
\frac{y_1 \sqrt{\rho c}}{\mathrm{i}}  \left( 1 + \frac{c^2 k^2}{ \beta^2} \right) + \frac{y_2}{\sqrt{\rho c}}\left(   \frac{2 c^2 \rho}{r\beta} - \frac{c^2 \rho }{\beta^2}\frac{\mathrm{d} \gamma}{\mathrm{d}z}  - \frac{1}{\beta}\frac{\mathrm{d} p}{\mathrm{d}z} \right) \nonumber \\ + \frac{c^2 \rho}{\beta} \frac{\mathrm{d}}{\mathrm{d}z} \left(\frac{y_2}{\sqrt{\rho c}} \right)=0
\label{y2s}
\end{eqnarray}
and
\begin{eqnarray}
\frac{g k^2}{ \beta^2}\frac{y_1 \sqrt{\rho c}}{\mathrm{i}}   + \frac{y_2}{\sqrt{\rho c}}\left( \rho \beta + \frac{2 g \rho}{r \beta} + \frac{g}{\beta} \frac{\mathrm{d} \rho}{\mathrm{d}z}- \frac{\rho }{\beta^2}\frac{\mathrm{d}\gamma}{\mathrm{d}z} \right) \nonumber \\ + \frac{g \rho}{\beta}\frac{\mathrm{d}}{\mathrm{d}z}\left(\frac{y_2}{\sqrt{\rho c}}\right) + \frac{\mathrm{d}}{\mathrm{d}z}\left(\frac{y_1 \sqrt{\rho c}}{\mathrm{i}}  \right)=0.
\label{y1s}
\end{eqnarray}
Then multiplying Equation~(\ref{y2s}) by $\sqrt{\rho c}$ and Equation~(\ref{y1s}) by $\mathrm{i} / \sqrt{\rho c}$ and rearranging,  we get
\begin{equation}
\frac{\mathrm{d} y_2}{\mathrm{d} z}=y_2\left( \frac{1}{\beta}\frac{\mathrm{d}\gamma}{\mathrm{d}z} - \frac{1}{2H_c} -  \frac{1}{2H_\rho}  -\frac{1}{\rho c^2}\frac{\mathrm{d} p}{\mathrm{d}z}\right) + \mathrm{i}  y_1 \left( \frac{\beta}{c}+ \frac{ck^2}{  \beta} \right)
\label{eighteen}
\end{equation}
and
\begin{equation}
\frac{\mathrm{d} y_1}{\mathrm{d}z}= - y_1 \left( \frac{1}{2 H_c} +\frac{1}{2 H_\rho} + \frac{g }{c^2} \right) + \mathrm{i}   y_2 \left( \frac{2g}{r \beta c} + \frac{g}{\beta c^3 \rho}\frac{\mathrm{d}p}{\mathrm{d}z} + \frac{g}{\beta \rho c}\frac{\mathrm{d} \rho}{\mathrm{d}z} - \frac{\beta}{c}   \right)
\label{nineteen}
\end{equation}
where $1/H_c =  -\mathrm{d}_z c/c$ and  $1/H_\rho=-\mathrm{d}_z \rho/\rho$. 
Equation~(\ref{eighteen}) and (\ref{nineteen}) reduce to Equations (A10) and (A11)   in  \cite{Birch2004} in the case where the attenuation is not dependent on $z$, the background is in hydrostatic equilibrium and the geometry is Cartesian.

The top boundary condition is a free surface such that the Lagrangian pressure perturbation [$\delta p$] is zero. This means that $p^\prime= - \bxi \cdot \bnabla p$. The bottom boundary is specified by $v_z=0$ and $p^\prime=1$. The boundary conditions translated to $y_1$ and $y_2$ are that $\rho c y_1 + \mathrm{i}  y_2 \beta \mathrm{d}_z p = 0$ at the top and $y_2=0$ and $y_1=1$ at the bottom.

We solve this boundary value problem using the Matlab program \texttt{bvp4c}. 
In order to be consistent with the eigenfunction solutions from the \textsf{SLiM} simulations, we do a similar normalisation of the eigenfunctions so that  $v_z(k,z=200~\mathrm{km})=1$.

\section{Solutions to the \textsf{BVP} for Different Background Models}\label{app3}
\begin{figure}
\hspace{1cm}
\includegraphics[width=0.9\textwidth]{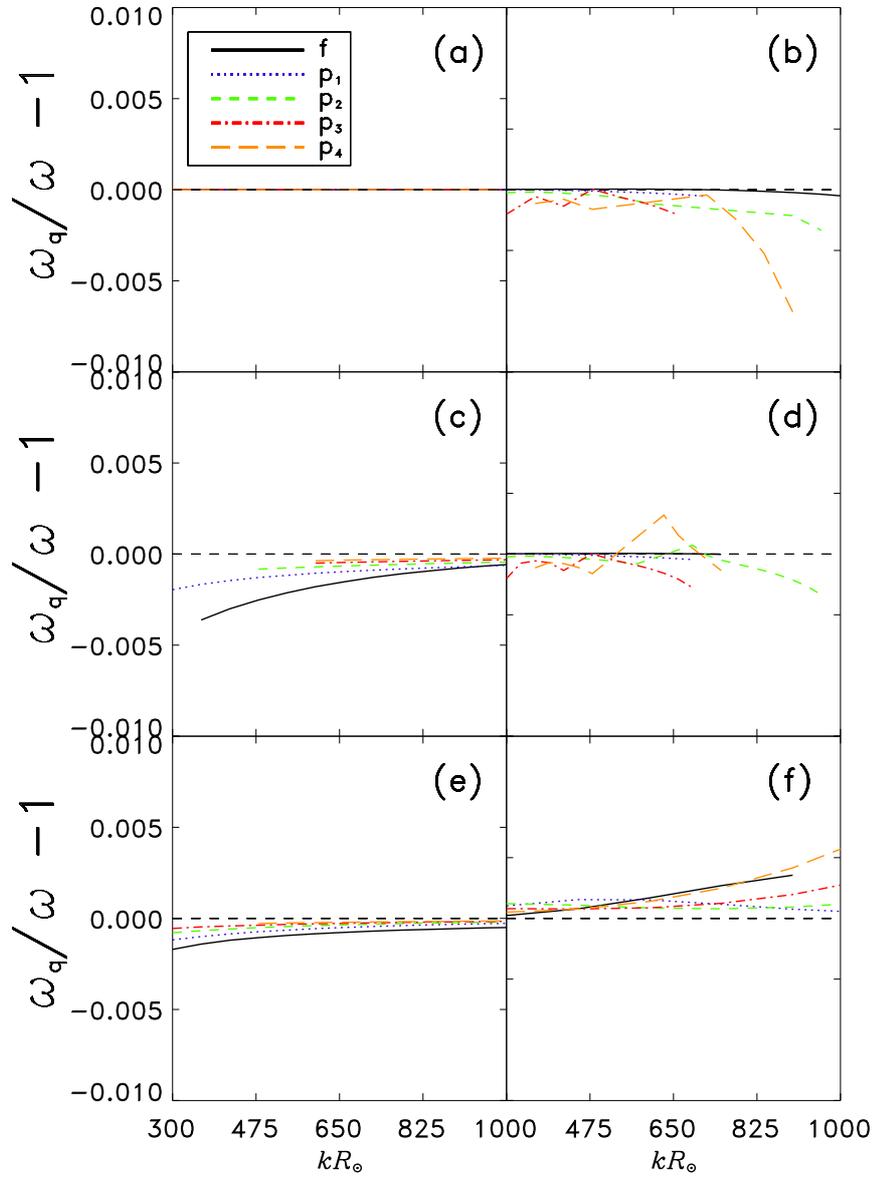}
\vspace{1cm}
\caption{The relative difference between the  CSM\_A eigenfrequencies with various modified quantities, $\omega_q$,  and the \textsf{BVP} eigenfrequencies of the original CSM\_A, $\omega$. The panels show the relative difference for with (a) 1\% noise added to the eigenfrequency guess, (b) no damping layers or attenuation, (c) constant gravity, (d) no damping layers, but retaining the attenuation, (e) Cartesian geometry and (f) top sponge extended lower in height. The frequency shifts are much smaller than those introduced by the convectively stable models.}
\label{bvpfig}
\end{figure}

We use the \textsf{BVP} solver outlined in Appendix~\ref{app2} to explore the effects on the eigenfrequencies by changing different parameters of the problem with CSM\_A.
To test the robustness of the \textsf{BVP} solver we added 1\% noise to the eigenfrequency guess that results in a relative difference of less than $10^{-5}$ as shown in  Figure~\ref{bvpfig} (a). 
In  Figure~\ref{bvpfig} (b) we do not apply any wave attenuation, \textit{i.e.} $\gamma = 0$. The eigenfrequencies decrease in value compared to the CSM\_A eigenfrequencies, \rev{more so} for the higher order modes.
In Figure~\ref{bvpfig} (c) we have set a constant gravitational acceleration of $g=-273.98 \, \mathrm{m/s^2}$. This mostly affects the \textit{f}-mode, but the eigenfrequencies are also decreased for the \textit{p}-modes.
Removing the sponge layers,  so that $\gamma = \Gamma(k)$, give results, Figure~\ref{bvpfig} (d), that are similar to  (b). 
Using the full Cartesian operators, \rev{as opposed to the spherical derivative in the radial direction as in Equation~\ref{eqn:A2}}, affects the eigenfrequencies at low-wavenumber the greatest, as shown in Figure~\ref{bvpfig} (e).
In Figure~\ref{bvpfig} (f) we have lowered the \textit{top} damping layer to have \rev{$\gamma(k,z)/2\pi = \Gamma(k)/4\pi + e^{ [(z+1.28~\mathrm{Mm}) / 0.25~\mathrm{Mm}] }\mu$Hz for $0.125<z<2.5$~Mm (retaining the bottom damping layer), which decreases the eigenfrequencies}. 
These frequency shifts are small compared to the frequency shifts caused by the convectively stabilising the models.

\clearpage

\acknowledgements
This work is supported by ERC grant agreement 210949, ``Seismic Imaging of the Solar Interior", to PI L. Gizon (Milestone \#4). We thank Aaron Birch for providing a set of Model~S eigenmodes, Shravan Hanasoge for the \textsf{SPARC} code, and Cristina Rabello-Soares for the full-disk MDI point spread function. SOHO is a mission of international collaboration between ESA and NASA.

\bibliographystyle{spr-mp-sola}
\bibliography{CSM}

\begin{thebibliography}{25}
\ifx \bisbn   \undefined \def \bisbn  #1{ISBN #1}\fi
\ifx \binits  \undefined \def \binits#1{#1}\fi
\ifx \bauthor  \undefined \def \bauthor#1{#1}\fi
\ifx \batitle  \undefined \def \batitle#1{#1}\fi
\ifx \bjtitle  \undefined \def \bjtitle#1{\textit{#1}}\fi
\ifx \bvolume  \undefined \def \bvolume#1{\textbf{#1}}\fi
\ifx \byear  \undefined \def \byear#1{#1}\fi
\ifx \bissue  \undefined \def \bissue#1{#1}\fi
\ifx \bfpage  \undefined \def \bfpage#1{#1}\fi
\ifx \blpage  \undefined \def \blpage #1{#1}\fi
\ifx \burl  \undefined \def \burl#1{\textsf{#1}}\fi
\ifx \href  \undefined \def \href#1#2{\textsf{#2}}\fi
\ifx \doiurl  \undefined \def
  \doiurl#1{\href{http://dx.doi.org/#1}{\textsf{#1}}}\fi
\ifx \betal  \undefined \def \betal{\textit{et al.}}\fi
\ifx \binstitute  \undefined \def \binstitute#1{#1}\fi
\ifx \bctitle  \undefined \def \bctitle#1{#1}\fi
\ifx \beditor  \undefined \def \beditor#1{#1}\fi
\ifx \bpublisher  \undefined \def \bpublisher#1{#1}\fi
\ifx \bbtitle  \undefined \def \bbtitle#1{\textit{#1}}\fi
\ifx \bedition  \undefined \def \bedition#1{#1}\fi
\ifx \bseriesno  \undefined \def \bseriesno#1{\textbf{#1}}\fi
\ifx \blocation  \undefined \def \blocation#1{#1}\fi
\ifx \bsertitle  \undefined \def \bsertitle#1{\textit{#1}}\fi
\ifx \bsnm \undefined \def \bsnm#1{#1}\fi
\ifx \bsuffix \undefined \def \bsuffix#1{#1}\fi
\ifx \bparticle \undefined \def \bparticle#1{#1}\fi
\ifx \barticle \undefined \def \barticle{}\fi
\ifx \botherref \undefined \def \botherref{}\fi
\ifx \url \undefined \def \url#1{\textsf{#1}}\fi
\ifx \bchapter \undefined \def \bchapter{}\fi
\ifx \bbook \undefined \def \bbook{}\fi
\ifx \bcomment \undefined \def \bcomment#1{#1}\fi
\ifx \oauthor \undefined \def \oauthor#1{#1}\fi
\ifx \citeauthoryear \undefined \def \citeauthoryear#1{#1}\fi
\def \endbibitem {}

\bibitem[\protect\citeauthoryear{{Antia} and {Basu}}{1999}]{Antia1999}
\begin{barticle}
\bauthor{\bsnm{{Antia}}, \binits{H.M.}}, \bauthor{\bsnm{{Basu}}, \binits{S.}}:
\byear{1999},
\batitle{{High-Frequency and High-Wavenumber Solar Oscillations}}.
\bjtitle{\apj}
\bvolume{519},
\bfpage{400}\,--\,\blpage{406}.
doi:\doiurl{10.1086/307364}.
\end{barticle}
\endbibitem

\bibitem[\protect\citeauthoryear{{Birch}, {Kosovichev}, and
  {Duvall}}{2004}]{Birch2004}
\begin{barticle}
\bauthor{\bsnm{{Birch}}, \binits{A.C.}}, \bauthor{\bsnm{{Kosovichev}},
  \binits{A.G.}}, \bauthor{\bsnm{{Duvall}}, \binits{T.L.} \bsuffix{Jr.}}:
\byear{2004},
\batitle{{Sensitivity of Acoustic Wave Travel Times to Sound-Speed
  Perturbations in the Solar Interior}}.
\bjtitle{\apj}
\bvolume{608},
\bfpage{580}\,--\,\blpage{600}.
doi:\doiurl{10.1086/386361}.
\end{barticle}
\endbibitem

\bibitem[\protect\citeauthoryear{{Bruls}}{1993}]{Bruls1993}
\begin{barticle}
\bauthor{\bsnm{{Bruls}}, \binits{J.H.M.J.}}:
\byear{1993},
\batitle{{The formation of helioseismology lines. IV - The NI I 676.8 NM
  intercombination line}}.
\bjtitle{\aap}
\bvolume{269},
\bfpage{509}\,--\,\blpage{517}.
\end{barticle}
\endbibitem

\bibitem[\protect\citeauthoryear{{Cally} and {Bogdan}}{1993}]{Cally1993}
\begin{barticle}
\bauthor{\bsnm{{Cally}}, \binits{P.S.}}, \bauthor{\bsnm{{Bogdan}},
  \binits{T.J.}}:
\byear{1993},
\batitle{{Solar p-modes in a vertical magnetic field - Trapped and damped
  pi-modes}}.
\bjtitle{\apj}
\bvolume{402},
\bfpage{721}\,--\,\blpage{732}.
doi:\doiurl{10.1086/172172}.
\end{barticle}
\endbibitem

\bibitem[\protect\citeauthoryear{{Cameron}, {Gizon}, and
  {Daiffallah}}{2007}]{Cameron2007}
\begin{barticle}
\bauthor{\bsnm{{Cameron}}, \binits{R.}}, \bauthor{\bsnm{{Gizon}}, \binits{L.}},
  \bauthor{\bsnm{{Daiffallah}}, \binits{K.}}:
\byear{2007},
\batitle{{SLiM: a code for the simulation of wave propagation through an
  inhomogeneous, magnetised solar atmosphere}}.
\bjtitle{Astronom. Nach.}
\bvolume{328},
\bfpage{313}.
doi:\doiurl{10.1002/asna.200610736}.
\end{barticle}
\endbibitem

\bibitem[\protect\citeauthoryear{{Cameron}, {Gizon}, and
  {Duvall}}{2008}]{Cameron2008}
\begin{barticle}
\bauthor{\bsnm{{Cameron}}, \binits{R.}}, \bauthor{\bsnm{{Gizon}}, \binits{L.}},
  \bauthor{\bsnm{{Duvall}}, \binits{T.L.} \bsuffix{Jr.}}:
\byear{2008},
\batitle{{Helioseismology of Sunspots: Confronting Observations with
  Three-Dimensional MHD Simulations of Wave Propagation}}.
\bjtitle{\solphys}
\bvolume{251},
\bfpage{291}\,--\,\blpage{308}.
doi:\doiurl{10.1007/s11207-008-9148-1}.
\end{barticle}
\endbibitem

\bibitem[\protect\citeauthoryear{Cameron \textit{et~al.}}{2011}]{Cameron2010}
\begin{barticle}
\bauthor{\bsnm{Cameron}, \binits{R.}}, \bauthor{\bsnm{Schunker}, \binits{H.}},
  \bauthor{\bsnm{Gizon}, \binits{L.}}, \bauthor{\bsnm{Pietarila}, \binits{A.}}:
\byear{2011},
\batitle{{Semi-empirical sunspot models for helioseismology}}.
\bjtitle{\solphys}
\bvolume{268},
\bfpage{293}\,--\,\blpage{308}.
doi:\doiurl{10.1007/s11207-010-9631-3}.
\end{barticle}
\endbibitem

\bibitem[\protect\citeauthoryear{{Christensen-Dalsgaard}
  \textit{et~al.}}{1996}]{JCD1996}
\begin{barticle}
\bauthor{\bsnm{{Christensen-Dalsgaard}}, \binits{J.}},
  \bauthor{\bsnm{{Dappen}}, \binits{W.}}, \bauthor{\bsnm{{Ajukov}},
  \binits{S.V.}}, \bauthor{\bsnm{{Anderson}}, \binits{E.R.}},
  \bauthor{\bsnm{{Antia}}, \binits{H.M.}}, \bauthor{\bsnm{{Basu}},
  \binits{S.}}, \bauthor{\bsnm{{Baturin}}, \binits{V.A.}},
  \bauthor{\bsnm{{Berthomieu}}, \binits{G.}}, \bauthor{\bsnm{{Chaboyer}},
  \binits{B.}}, \bauthor{\bsnm{{Chitre}}, \binits{S.M.}},
  \bauthor{\bsnm{{Cox}}, \binits{A.N.}}, \bauthor{\bsnm{{Demarque}},
  \binits{P.}}, \bauthor{\bsnm{{Donatowicz}}, \binits{J.}},
  \bauthor{\bsnm{{Dziembowski}}, \binits{W.A.}}, \bauthor{\bsnm{{Gabriel}},
  \binits{M.}}, \bauthor{\bsnm{{Gough}}, \binits{D.O.}},
  \bauthor{\bsnm{{Guenther}}, \binits{D.B.}}, \bauthor{\bsnm{{Guzik}},
  \binits{J.A.}}, \bauthor{\bsnm{{Harvey}}, \binits{J.W.}},
  \bauthor{\bsnm{{Hill}}, \binits{F.}}, \bauthor{\bsnm{{Houdek}}, \binits{G.}},
  \bauthor{\bsnm{{Iglesias}}, \binits{C.A.}}, \bauthor{\bsnm{{Kosovichev}},
  \binits{A.G.}}, \bauthor{\bsnm{{Leibacher}}, \binits{J.W.}},
  \bauthor{\bsnm{{Morel}}, \binits{P.}}, \bauthor{\bsnm{{Proffitt}},
  \binits{C.R.}}, \bauthor{\bsnm{{Provost}}, \binits{J.}},
  \bauthor{\bsnm{{Reiter}}, \binits{J.}}, \bauthor{\bsnm{{Rhodes}},
  \binits{E.J.} \bsuffix{Jr.}}, \bauthor{\bsnm{{Rogers}}, \binits{F.J.}},
  \bauthor{\bsnm{{Roxburgh}}, \binits{I.W.}}, \bauthor{\bsnm{{Thompson}},
  \binits{M.J.}}, \bauthor{\bsnm{{Ulrich}}, \binits{R.K.}}:
\byear{1996},
\batitle{{The Current State of Solar Modeling}}.
\bjtitle{Science}
\bvolume{272},
\bfpage{1286}\,--\,\blpage{1292}.
\end{barticle}
\endbibitem

\bibitem[\protect\citeauthoryear{{Dahlen} and {Tromp}}{1998}]{Dahlen1998}
\begin{bbook}
\bauthor{\bsnm{{Dahlen}}, \binits{F.A.}}, \bauthor{\bsnm{{Tromp}},
  \binits{J.}}:
\byear{1998},
\bbtitle{Theoretical global seismology},
\bpublisher{Princeton University Press},
\blocation{Princeton, New Jersey},
\bfpage{120}.
\end{bbook}
\endbibitem

\bibitem[\protect\citeauthoryear{Dombroski, Birch, and
  Braun}{2011}]{Dombroski2011}
\begin{botherref}
\oauthor{\bsnm{Dombroski}, \binits{D.}}, \oauthor{\bsnm{Birch}, \binits{A.}},
  \oauthor{\bsnm{Braun}, \binits{D.}}:
2011,
{Testing Helioseismic Holography Inversions for Supergranular Flows Using
  Synthetic Data}.
\textit{\solphys, in prep}.
\end{botherref}
\endbibitem

\bibitem[\protect\citeauthoryear{{Duvall} \textit{et~al.}}{1993}]{Duvall1993}
\begin{barticle}
\bauthor{\bsnm{{Duvall}}, \binits{T.L.} \bsuffix{Jr.}},
  \bauthor{\bsnm{{Jefferies}}, \binits{S.M.}}, \bauthor{\bsnm{{Harvey}},
  \binits{J.W.}}, \bauthor{\bsnm{{Osaki}}, \binits{Y.}},
  \bauthor{\bsnm{{Pomerantz}}, \binits{M.A.}}:
\byear{1993},
\batitle{{Asymmetries of solar oscillation line profiles}}.
\bjtitle{\apj}
\bvolume{410},
\bfpage{829}\,--\,\blpage{836}.
doi:\doiurl{10.1086/172800}.
\end{barticle}
\endbibitem

\bibitem[\protect\citeauthoryear{{Gizon}}{2006}]{Gizon2006}
\begin{barticle}
\bauthor{\bsnm{{Gizon}}, \binits{L.}}:
\byear{2006},
\batitle{{Line Profiles of Fundamental Modes of Solar Oscillation}}.
\bjtitle{Central Euro. Astrophys. Bull.}
\bvolume{30},
\bfpage{1}\,--\,\blpage{9}.
\end{barticle}
\endbibitem

\bibitem[\protect\citeauthoryear{{Gizon} and {Birch}}{2002}]{Gizon2002}
\begin{barticle}
\bauthor{\bsnm{{Gizon}}, \binits{L.}}, \bauthor{\bsnm{{Birch}}, \binits{A.C.}}:
\byear{2002},
\batitle{{Time-Distance Helioseismology: The Forward Problem for Random
  Distributed Sources}}.
\bjtitle{\apj}
\bvolume{571},
\bfpage{966}\,--\,\blpage{986}.
doi:\doiurl{10.1086/340015}.
\end{barticle}
\endbibitem

\bibitem[\protect\citeauthoryear{{Gizon} and {Birch}}{2004}]{Gizon2004}
\begin{barticle}
\bauthor{\bsnm{{Gizon}}, \binits{L.}}, \bauthor{\bsnm{{Birch}}, \binits{A.C.}}:
\byear{2004},
\batitle{{Time-Distance Helioseismology: Noise Estimation}}.
\bjtitle{\apj}
\bvolume{614},
\bfpage{472}\,--\,\blpage{489}.
doi:\doiurl{10.1086/423367}.
\end{barticle}
\endbibitem

\bibitem[\protect\citeauthoryear{{Hanasoge}, {Duvall}, and
  {Couvidat}}{2007}]{Hanasoge2007}
\begin{barticle}
\bauthor{\bsnm{{Hanasoge}}, \binits{S.M.}}, \bauthor{\bsnm{{Duvall}},
  \binits{T.L.} \bsuffix{Jr.}}, \bauthor{\bsnm{{Couvidat}}, \binits{S.}}:
\byear{2007},
\batitle{{Validation of Helioseismology through Forward Modeling: Realization
  Noise Subtraction and Kernels}}.
\bjtitle{\apj}
\bvolume{664},
\bfpage{1234}\,--\,\blpage{1243}.
doi:\doiurl{10.1086/519070}.
\end{barticle}
\endbibitem

\bibitem[\protect\citeauthoryear{{Hanasoge}
  \textit{et~al.}}{2006}]{Hanasoge2006}
\begin{barticle}
\bauthor{\bsnm{{Hanasoge}}, \binits{S.M.}}, \bauthor{\bsnm{{Larsen}},
  \binits{R.M.}}, \bauthor{\bsnm{{Duvall}}, \binits{T.L.} \bsuffix{Jr.}},
  \bauthor{\bsnm{{De Rosa}}, \binits{M.L.}}, \bauthor{\bsnm{{Hurlburt}},
  \binits{N.E.}}, \bauthor{\bsnm{{Schou}}, \binits{J.}},
  \bauthor{\bsnm{{Roth}}, \binits{M.}},
  \bauthor{\bsnm{{Christensen-Dalsgaard}}, \binits{J.}},
  \bauthor{\bsnm{{Lele}}, \binits{S.K.}}:
\byear{2006},
\batitle{{Computational Acoustics in Spherical Geometry: Steps toward
  Validating Helioseismology}}.
\bjtitle{\apj}
\bvolume{648},
\bfpage{1268}\,--\,\blpage{1275}.
doi:\doiurl{10.1086/505927}.
\end{barticle}
\endbibitem

\bibitem[\protect\citeauthoryear{{Lynden-Bell} and
  {Ostriker}}{1967}]{LyndenBell1967}
\begin{barticle}
\bauthor{\bsnm{{Lynden-Bell}}, \binits{D.}}, \bauthor{\bsnm{{Ostriker}},
  \binits{J.P.}}:
\byear{1967},
\batitle{{On the stability of differentially rotating bodies}}.
\bjtitle{\mnras}
\bvolume{136},
\bfpage{293}.
\end{barticle}
\endbibitem

\bibitem[\protect\citeauthoryear{{Nigam} and {Kosovichev}}{1999}]{Nigam1999}
\begin{barticle}
\bauthor{\bsnm{{Nigam}}, \binits{R.}}, \bauthor{\bsnm{{Kosovichev}},
  \binits{A.G.}}:
\byear{1999},
\batitle{{Source of Solar Acoustic Modes}}.
\bjtitle{\apjl}
\bvolume{514},
\bfpage{L53}\,--\,\blpage{L56}.
doi:\doiurl{10.1086/311939}.
\end{barticle}
\endbibitem

\bibitem[\protect\citeauthoryear{{Parchevsky} and
  {Kosovichev}}{2007}]{Parchevsky2007}
\begin{barticle}
\bauthor{\bsnm{{Parchevsky}}, \binits{K.V.}}, \bauthor{\bsnm{{Kosovichev}},
  \binits{A.G.}}:
\byear{2007},
\batitle{{Three-dimensional Numerical Simulations of the Acoustic Wave Field in
  the Upper Convection Zone of the Sun}}.
\bjtitle{\apj}
\bvolume{666},
\bfpage{547}\,--\,\blpage{558}.
doi:\doiurl{10.1086/520108}.
\end{barticle}
\endbibitem

\bibitem[\protect\citeauthoryear{{Rabello-Soares}, {Korzennik}, and
  {Schou}}{2001}]{RabelloSoares2001}
\begin{bchapter}
\bauthor{\bsnm{{Rabello-Soares}}, \binits{M.C.}}, \bauthor{\bsnm{{Korzennik}},
  \binits{S.G.}}, \bauthor{\bsnm{{Schou}}, \binits{J.}}:
\byear{2001},
\bctitle{{The determination of MDI high-degree mode frequencies}}.
In: \beditor{\bsnm{Pall{\'e}}, \binits{W..}} (ed.)
\bbtitle{SOHO 10/GONG 2000 Workshop: Helio- and Asteroseismology at the Dawn of
  the Millennium}
\bseriesno{SP-464},
\bpublisher{ESA},
\blocation{Nordwijk},
\bfpage{129}\,--\,\blpage{136}.
\end{bchapter}
\endbibitem

\bibitem[\protect\citeauthoryear{{Scherrer}
  \textit{et~al.}}{1995}]{Scherrer1995}
\begin{barticle}
\bauthor{\bsnm{{Scherrer}}, \binits{P.H.}}, \bauthor{\bsnm{{Bogart}},
  \binits{R.S.}}, \bauthor{\bsnm{{Bush}}, \binits{R.I.}},
  \bauthor{\bsnm{{Hoeksema}}, \binits{J.T.}}, \bauthor{\bsnm{{Kosovichev}},
  \binits{A.G.}}, \bauthor{\bsnm{{Schou}}, \binits{J.}},
  \bauthor{\bsnm{{Rosenberg}}, \binits{W.}}, \bauthor{\bsnm{{Springer}},
  \binits{L.}}, \bauthor{\bsnm{{Tarbell}}, \binits{T.D.}},
  \bauthor{\bsnm{{Title}}, \binits{A.}}, \bauthor{\bsnm{{Wolfson}},
  \binits{C.J.}}, \bauthor{\bsnm{{Zayer}}, \binits{I.}}, \bauthor{\bsnm{{MDI
  Engineering Team}}}:
\byear{1995},
\batitle{{The Solar Oscillations Investigation - Michelson Doppler Imager}}.
\bjtitle{\solphys}
\bvolume{162},
\bfpage{129}\,--\,\blpage{188}.
doi:\doiurl{10.1007/BF00733429}.
\end{barticle}
\endbibitem

\bibitem[\protect\citeauthoryear{{Schunker} and {Gizon}}{2008}]{Schunker2008}
\begin{barticle}
\bauthor{\bsnm{{Schunker}}, \binits{H.}}, \bauthor{\bsnm{{Gizon}},
  \binits{L.}}:
\byear{2008},
\batitle{{HELAS Local Helioseismology Activities}}.
\bjtitle{Comm. in Asteroseis.}
\bvolume{156},
\bfpage{93}\,--\,\blpage{105}.
\end{barticle}
\endbibitem

\bibitem[\protect\citeauthoryear{{Schunker}, {Cameron}, and
  {Gizon}}{2010}]{Schunker2010}
\begin{botherref}
\oauthor{\bsnm{{Schunker}}, \binits{H.}}, \oauthor{\bsnm{{Cameron}},
  \binits{R.}}, \oauthor{\bsnm{{Gizon}}, \binits{L.}}:
2010,
{Convectively stabilised background solar models for local helioseismology}.
\textit{ArXiv e-prints: 1002.1969}.
\end{botherref}
\endbibitem

\bibitem[\protect\citeauthoryear{{Shelyag}, {Fedun}, and
  {Erd{\'e}lyi}}{2008}]{Shelyag2008}
\begin{barticle}
\bauthor{\bsnm{{Shelyag}}, \binits{S.}}, \bauthor{\bsnm{{Fedun}}, \binits{V.}},
  \bauthor{\bsnm{{Erd{\'e}lyi}}, \binits{R.}}:
\byear{2008},
\batitle{{Magnetohydrodynamic code for gravitationally-stratified media}}.
\bjtitle{\aap}
\bvolume{486},
\bfpage{655}\,--\,\blpage{662}.
doi:\doiurl{10.1051/0004-6361:200809800}.
\end{barticle}
\endbibitem

\bibitem[\protect\citeauthoryear{{Title} \textit{et~al.}}{1989}]{Title1989}
\begin{barticle}
\bauthor{\bsnm{{Title}}, \binits{A.M.}}, \bauthor{\bsnm{{Tarbell}},
  \binits{T.D.}}, \bauthor{\bsnm{{Topka}}, \binits{K.P.}},
  \bauthor{\bsnm{{Ferguson}}, \binits{S.H.}}, \bauthor{\bsnm{{Shine}},
  \binits{R.A.}}, \bauthor{\bsnm{{SOUP Team}}}:
\byear{1989},
\batitle{{Statistical properties of solar granulation derived from the SOUP
  instrument on Spacelab 2}}.
\bjtitle{\apj}
\bvolume{336},
\bfpage{475}\,--\,\blpage{494}.
doi:\doiurl{10.1086/167026}.
\end{barticle}
\endbibitem

\end{thebibliography}

\end{document}